\documentclass{aa}  %este es la version comun a dos cols.

\usepackage{graphicx}
\usepackage{longtable}
\usepackage{txfonts}
\begin{document}

   \title{A temperature condensation trend in the debris-disk binary system $\zeta^{2}$ Ret
   %\thanks{The data presented herein were obtained at the W.M. Keck Observatory, which is operated as a
   %scientific partnership among the California Institute of Technology, the University of California, and the
   %National Aeronautics and Space Administration. The Observatory was made possible by the generous
   %financial support of the W.M. Keck Foundation.}
   %\footnote{Reduced spectra of HD 80606 and HD 80607 (FITS files) are only available in electronic
   %form at the CDS via anonymous ftp to cdsarc.u-strasbg.fr (130.79.128.5) or
   %via http://cdsweb.u-strasbg.fr/cgi-bin/qcat?J/A+A/}
   }

   %\subtitle{I. Overviewing the $\kappa$-mechanism}
   
   \titlerunning{A temperature condensation trend in $\zeta^{2}$ Ret}
   \authorrunning{Saffe et al.}

   \author{C. Saffe\inst{1,2,6}, M. Flores\inst{1,6}, M. Jaque Arancibia\inst{1,6}, A. Buccino\inst{3,4,6} \and E. Jofr\'e\inst{5,6}
   %\fnmsep\thanks{Just to show the usage of the elements in the author field}
          }

   \institute{Instituto de Ciencias Astron\'omicas, de la Tierra y del Espacio (ICATE-CONICET),
              C.C 467, 5400, San Juan, Argentina.
             \email{csaffe,mflores,mjaque@icate-conicet.gob.ar}
         \and
         Universidad Nacional de San Juan (UNSJ), Facultad de Ciencias Exactas, F\'isicas y Naturales (FCEFN), San Juan, Argentina.
         \and
         Instituto de Astronom\'ia y F\'isica del Espacio (IAFE-CONICET), Buenos Aires, Argentina.
             \email{abuccino@iafe.uba.ar}
         \and
         Departamento de F\'isica, Facultad de Ciencias Exactas y Naturales (FCEN), Universidad de Buenos Aires (UBA), Buenos Aires, Argentina. 
         \and
         Observatorio Astron\'omico de C\'ordoba (OAC), Laprida 854, X5000BGR, C\'ordoba, Argentina.
             \email{emiliano@oac.uncor.edu}
         \and
         Consejo Nacional de Investigaciones Cient\'ificas y T\'ecnicas (CONICET), Argentina         
        }

   \date{Received xxx, xxx ; accepted xxxx, xxxx}

% \abstract{}{}{}{}{} 
% 5 {} token are mandatory
 
  \abstract
   {Detailed abundance studies have reported different trends between samples of stars
   with and without planets, possibly related to the planet formation process.
   Whether these differences are still present between samples of stars with and without
   debris disk is still unclear.}
   {We explore condensation temperature T$_{c}$ trends in the unique binary system $\zeta^{1}$ Ret -
   $\zeta^{2}$ Ret, to determine whether there is a depletion of refractories, which could be
   related to the planet formation process.
   The star $\zeta^{2}$ Ret hosts a debris disk which was detected by an IR excess and confirmed
   by direct imaging and numerical simulations, while $\zeta^{1}$ Ret does not present IR excess
   nor planets. These characteristics convert $\zeta^{2}$ Ret in a remarkable system,
   where their binary nature together with the strong similarity of both components allow us,
   for the first time, to achieve the highest possible abundance precision in this system.
   }
   {We carried out a high-precision abundance determination in both components of the binary system
   via a line-by-line, strictly differential approach. First, we used the Sun as a reference and
   then we used $\zeta^{2}$ Ret. The stellar parameters T$_{eff}$, {log $g$}, [Fe/H] and v$_{turb}$
   were determined by imposing differential ionization and excitation equilibrium of Fe I and Fe II
   lines, with an updated version of the program FUNDPAR, together with plane-parallel local
   thermodynamic equilibrium (LTE) ATLAS9 model atmospheres and the MOOG code.
   Then, we derived detailed abundances of 24 different species with equivalent widths and spectral
   synthesis with the MOOG program. The chemical patterns were compared with the solar-twins
   T$_{c}$ trend of \citet{melendez09}, and then mutually between both stars of the binary system.
   The rocky mass of depleted refractory material was estimated according to \citet{chambers10}.
   }
   {The star $\zeta^{1}$ Ret resulted slightly more metal rich than $\zeta^{2}$ Ret by $\sim$0.02 dex.
   In the differential calculation of $\zeta^{1}$ Ret using $\zeta^{2}$ Ret as reference, the
   abundances of the refractory elements resulted higher than the volatile elements, and the trend
   of the refractory elements with T$_{c}$ showed a positive slope.
   These facts together show a lack of refractory elements in $\zeta^{2}$ Ret (a debris-disk host)
   relative to $\zeta^{1}$ Ret. %This is contrary to the recent result of \citet{maldonado15}, probably due to the increased precision of the differential technique applied here.
   The T$_{c}$ trend would be in agreement with the proposed signature of planet formation
   \citep{melendez09} rather than possible Galactic Chemical Evolution or age effects, which are
   largely diminished here.
   Then, following the interpretation of \citet{melendez09}, we propose an scenario in which the
   refractory elements depleted in $\zeta^{2}$ Ret are possibly locked-up in the rocky material that
   orbits this star and produce the debris disk observed around this object.
   We estimated a lower limit of M$_{rock} \sim$ 3 M$_{\oplus}$ for the rocky mass of depleted material,
   which is compatible with a rough estimation of 3-50 M$_{\oplus}$ of a debris disk mass around a
   solar-type star \citep{krivov08}.}
   {}
   

   \maketitle
%
%________________________________________________________________

\section{Introduction}

 \citet[][ hereafter M09]{melendez09} showed that the Sun is deficient in refractory elements 
 relative to volatile elements when compared to the mean abundances of 11 solar twins. They also
 found that the abundance differences correlate strongly with the condensation temperature T$_{c}$,
 which is interpreted by the authors as a possible signature of terrestrial planet formation in the
 Solar System. They suggested that the refractory elements (T$_{c}>$ 900 K) depleted in the solar
 photosphere are locked up in terrestrial planets and rocky material at the time of star and
 planet formation. In a follow-up study, \citet[][ hereafter R10]{ramirez10} confirmed their findings.
 \citet{gonzalez10} and \citet{gonzalez11} also found that most metal-rich exoplanet host (EH)
 stars have the most negative trends. These results indicate that the depletion of refractory
 elements in a photosphere is a consequence of both terrestrial and giant planet formation.
 If this hypothesis is correct, stars with planetary systems like ours may be identified
 through a very detailed inspection of the chemical composition. 

 Debris disks orbiting main-sequence stars are observationally characterized by an infrared
 excess over the normal photospheric fluxes of their host stars
 \citep[see e.g.][]{aumann84,mannings-barlow98,habing01,bryden06,beichman06}.
 The excess is produced by the presence of dust in the disk which is attributed to the collisions
 of larger rocky bodies
 \citep[see e.g. the reviews of][and references therein]{wyatt08,moro-martin13,matthews14}.
 The existence of these dusty disks is confirmed in some cases by direct imaging, starting with
 the first $\beta$ Pictoris image by \citet{smith-terrile84} and then followed by other examples
 \citep[see e.g.][]{krist05,vanden10,soumer14,currie15}.
 \citet[][ hereafter MA15]{maldonado15} compared T$_{c}$ trends in a sample of stars with debris
 disks and stars with neither debris nor planets, and found no statistical difference between them. 
 In other words, they do not detect a possible lack of refractory elements in debris disk stars
 when compared to stars without disks. The detection of a T$_{c}$ trend is a challenging task that
 requires the highest possible precision in abundances, such as those obtained with the
 line-by-line differential technique \citep[e.g.][]{bedell14,saffe15}.
 However, the sample of MA15 included 251 FGK stars spanning a range in T$_{eff}$ of $\sim$ 2000 K, which
 prevented the authors to perform a differential analysis, as they explained.
 In addition, the study of binary systems with similar components greatly diminishes other T$_{c}$
 effects such as the Galactic Chemical Evolution \citep[GCE,][]{gonz-hern13,schuler11},
 the stellar age and a possible inner galactic origin of the planet hosts \citep[e.g.][]{adi14},
 thanks to the common natal environment of the pair.
 MA15 showed that these effects are indeed present in their sample, suggesting that an evolutionary
 effect is present.
 Then, the study of binary systems with similar components presents important advantages
 aiming to detect a possible T$_{c}$ trend related to the planet formation process
 \citep[e.g.][]{tucci-maia14,saffe15}.
 
 Up to now, to find a binary system (with similar components) where only one star hosts
 a debris disk is a very difficult task. Most of the IR surveys performed with the satellite
 Spitzer have been mainly focused on single main-sequence stars rather than multiple systems
 \citep[e.g.][]{beichman05,rieke05,bryden06,su06}.
 Using Spitzer data, the location of the dust in a multiple system (i.e. circumstellar, circumbinary
 or both) could be determined only in very few cases \citep[see e.g.][]{trilling07}.
 \citet{rodriguez-zuckerman12} showed that only $\sim$ 25\% of the debris disk stars in their
 sample of 112 main-sequence stars belong to multiple systems.
 The satellite Herschel overcome some of these difficulties, thanks to their greater sensivity
 and spatial resolution. Different Herschel surveys such as SUNS, DEBRIS \citep{matthews10,phillips10},
 and DUNES \citep{eiroa10,lohne12,eiroa13} do include multiple systems in their samples. However, the
 circumstellar nature of the dust have been determined in few multiple systems
 \citep[see e.g.][]{eiroa13}. Recently, \citet{rodriguez15} estimated a multiplicity of $\sim$ 42\%
 in the stars of the DEBRIS survey by using adaptive optics imaging. However, there is no physical
 data of many systems and most of them do not present similar components (see e.g. their Table 1).
 This shows how particular could be to find such kind of binary system.

 As a part of the DUNES survey, \citet{eiroa10} discovered a resolved debris disk around the
 star $\zeta^{2}$ Ret (= HD 20807), which is accompanied by the star $\zeta^{1}$ Ret (= HD 20766).
 The projected distance between the stars is 3713 AU \citep{mason01}, while a
 Bayesian analysis of the proper motions indicates a very high probability (near 100\%)
 that the pair is physically connected \citep{shaya-olling11}.
 This is an unique system for a number of reasons.
 The presence of a debris disk around $\zeta^{2}$ Ret was detected through a mid-IR excess \citep{trilling08,eiroa13},
 it was {\it{confirmed}} by direct imaging \citep{eiroa10}, and also supported by numerical simulations \citep{faramaz14}.
 On the other hand, the companion $\zeta^{1}$ Ret does not present IR excess using both Spitzer
 and Herschel observations \citep{bryden06,trilling08,eiroa13}.
 The spectral types of the binary components are very similar (G2 V + G1 V, as appear in the Hipparcos database)
 allowing a chemical comparison less dependent of the fundamental parameters of the stars.
 Both stars are also very similar to the Sun, being both solar-analogs.
 Also, there is no planet detected in this binary system (as we explain below), which is a condition
 for the sample of MA15.
 These characteristics shows that this is a remarkable binary system, an ideal case to test
 a possible T$_{c}$ trend in stars with and without debris disks.
 
 Using numerical simulations, \citet{faramaz14} suggested that the eccentric structure of
 the $\zeta^{2}$ Ret debris disk could be caused by an eccentric (e $>$ 0.3) planetary companion.
 However, both stars ($\zeta^{1}$ and $\zeta^{2}$) have been monitored by the Anglo-Australian Planet Search
 (AAPS) radial-velocity survey\footnote{http://newt.phys.unsw.edu.au/~cgt/planet/Targets.html},
 while $\zeta^{2}$ Ret was also included in the HARPS GTO planet search program \citep[e.g.][]{sousa08}
 giving no-planet detection.
 The AAPS survey allows to rule out a Saturn-mass (0.3 M$_{Jup}$) or a larger planet with a period
 range P < 300 days and eccentricity 0.0 < e < 0.6 \citep[]{wittenmyer10}.
 The HARPS GTO survey suggests that there is no Jupiter-mass or a larger planet interior to
 $\sim$ 5–10 AU \citep{mayor03}.
 Although the radial-velocity surveys cannot completely rule out the presence of planets
 (such as e.g. long period perturbers or lower-mass planets),
 these stars form, to our knowledge, the only solar-like binary system with similar
 components where only one component hosts a debris disk (confirmed by direct imaging) and
 the presence of planets has not been confirmed yet.
 
 Surprinsingly, in this work we found a T$_{c}$ trend between the components of this binary system i.e.
 a lack of refractory elements in a debris-disk star. 
 %This is contrary to the recent result of MA15, probably due to the greater sensivity of our differential technique.
 We note that the MA15 statistical result does not exclude that Tc trends may be present
 in particular stars, such as the components of the $\zeta$ Ret system.
 The T$_{c}$ trend would be in agreement with the proposed signature of planet formation
 \citep[e.g.][]{melendez09,ramirez10} rather than possible GCE, age or evolutionary effects
 which are largely diminished here.
 
 The abundances of the stars have been previously determined in the literature.
 However, there are some differences in the fundamental parameters derived for $\zeta^{2}$ Ret.
 The reported metallicities vary from -0.16 dex to -0.36 dex \citep{maldonado12,allende04}
 while log g vary between 4.41 dex and 4.64 dex \citep{bensby14,maldonado12}.
 These differences also encouraged this work, searching for a slight difference between the
 components of this binary system.

This work is organized as follows.
In Section §2 we describe the observations and data reduction, while in Section §3 we present
the stellar parameters and chemical abundance analysis. In Section §4 we show the results and
discussion, and finally in Section §5 we highlight our main conclusions.

\section{Observations and data reduction}

Stellar spectra of $\zeta^{1}$ Ret and $\zeta^{2}$ Ret were obtained with the
High Accuracy Radial velocity Planet Searcher (HARPS) spectrograph,
attached to the La Silla 3.6m (ESO) telescope.
The spectrograph was fed by a pair of fibres with an aperture of 1 arcsec
on the sky, resulting a resolving power of $\sim$
115000\footnote{http://www.eso.org/sci/facilities/lasilla/instruments/harps/overview.html}.
The spectra were obtained from the ESO HARPS
archive\footnote{https://www.eso.org/sci/facilities/lasilla/instruments/harps/tools/
archive.html},
under the ESO program identification 072.C-0513(D).

The observations were taken on February, 4th 2004 with $\zeta^{2}$ Ret observed
immediately after $\zeta^{1}$ Ret, using the same spectrograph configuration.
The exposure times were 3 x 150 s for both targets.
We measured a signal-to-noise S/N $\sim$ 300 for each of the binary components, with an
spectral covarge between 3870-6900 \AA.
The asteroid Ganymede was also observed with the same spectrograph set-up
achieving a slightly higher S/N, to acquire the solar spectrum useful for
reference in our (initial) differential analysis. We note however that
the final differential study with the highest abundance precision is between
$\zeta^{1}$ Ret and $\zeta^{2}$ Ret because of their high degree of similarity.
The data were reduced with the HARPS pipeline and combined using the software
package IRAF\footnote{IRAF is distributed by the National Optical Astronomical Observatories,
which is operated by the Association of Universities for Research in Astronomy, Inc. under
a cooperative agreement with the National Science Foundation.}.

\section{Stellar parameters and chemical abundance analysis}

We derived the fundamental parameters (T$_{eff}$, {log $g$}, [Fe/H], v$_{turb}$) of $\zeta^{1}$ Ret and
$\zeta^{2}$ Ret following the same procedure detailed in our previous work \citep{saffe15}.
We started by measuring the equivalent widths (EW) of Fe I and Fe II lines in the spectra
of our program stars using the IRAF task {\it{splot}}, and then continued with other
chemical species. The lines list and relevant laboratory data (such as excitation potential and
oscilator strengths) were taken from \citet{liu14}, \citet{melendez14}, and then extended
with data from \citet{bedell14}, who carefully selected lines for a high-precision abundance
determination. This data, including the measured EWs, are presented in Table \ref{linelist}.
Then, we imposed excitation and ionization balance of Fe I and Fe II lines, using the differential
version of the program FUNDPAR \citep{saffe11}, together with the 2014 version of the MOOG code
\citep{sneden73} and ATLAS9 model atmospheres \citep{kurucz93}.

Stellar parameters of $\zeta^{1}$ Ret and $\zeta^{2}$ Ret were differentially determined using the Sun
as standard in an initial approach, and then we recalculated the parameters of $\zeta^{1}$ Ret using 
$\zeta^{2}$ Ret as reference. First, we determined absolute abundances for the Sun using
{5777 K} for T$_{eff}$, {4.44 dex} for {log $g$} and an initial v$_{turb}$ of {1.0 km/s}. Then,
we estimated v$_{turb}$ for the Sun with the usual method of requiring zero slope in the
absolute abundances of {Fe I} lines versus EW$_{r}$ and obtained a final v$_{turb}$ of {0.91 km/s}.
We note however that the exact values are not crucial for our strictly differential study
\citep[see e.g.][]{bedell14,saffe15}.

\begin{figure}
\centering
\includegraphics[width=8cm]{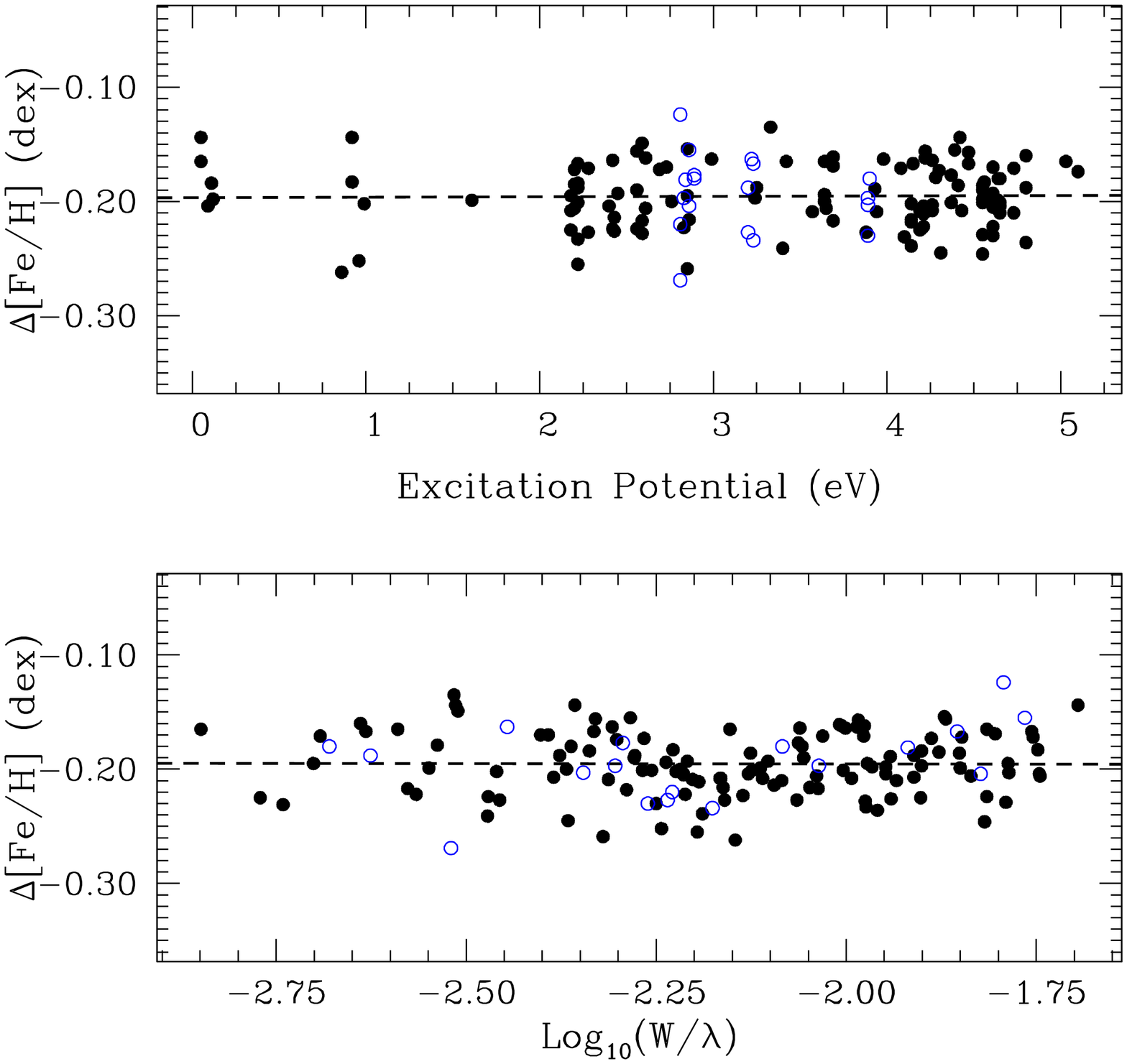}
\caption{Differential abundance vs excitation potential (upper panel) 
and differential abundance vs reduced EW (lower panel), for $\zeta^{1}$ Ret relative to the Sun.
Filled and empty points correspond to Fe I and Fe II, respectively.
The dashed line is a linear fit to the abundance values.}
\label{equil-zet01-sun}%
\end{figure}

\begin{figure}
\centering
\includegraphics[width=8cm]{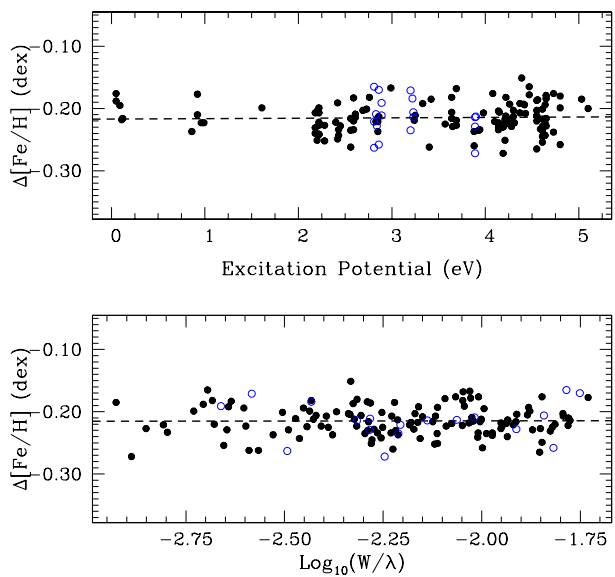}
\caption{Differential abundance vs excitation potential (upper panel) 
and differential abundance vs reduced EW (lower panel), for $\zeta^{2}$ Ret relative to the Sun.
Filled and empty points correspond to Fe I and Fe II, respectively.
The dashed line is a linear fit to the abundance values.}
\label{equil-zet02-sun}%
\end{figure}

The next step was the determination of stellar parameters of $\zeta^{1}$ Ret and $\zeta^{2}$ Ret using the
Sun as standard, i.e. ($\zeta^{1}$ Ret - Sun) and ($\zeta^{2}$ Ret - Sun).
For $\zeta^{1}$ Ret the resulting stellar parameters were {T$_{eff}$ = 5710$\pm$29 K},
{log $g$ = 4.53$\pm$0.05 dex}, {[Fe/H] = -0.195$\pm$0.005 dex,} and {v$_{turb}$ = 0.80$\pm$0.27 km/s}.
For $\zeta^{2}$ Ret, we obtained {T$_{eff}$ = 5854$\pm$28 K},
{log $g$ = 4.54$\pm$0.04 dex}, {[Fe/H] = -0.215$\pm$0.004 dex,} and {v$_{turb}$ = 0.95$\pm$0.09 km/s}.
In the Table \ref{param.dif} we present the differential parameters $\Delta$T$_{eff}$, $\Delta$log $g$,
$\Delta$(Fe/H) and $\Delta$v$_{turb}$ derived between the star and their corresponding reference.
The errors in the stellar parameters were derived following the procedure detailed in \citet{saffe15},
which takes into account the individual and the mutual covariance terms of the error propagation.
The star $\zeta^{1}$ Ret resulted with a slightly higher metallicity than $\zeta^{2}$ Ret by $\sim$ 0.02 dex.
Figures \ref{equil-zet01-sun} and \ref{equil-zet02-sun} show abundance vs excitation
potential and abundance vs EW$_{r}$ for both stars. Filled and empty points correspond to Fe I and
Fe II, while the dashed lines are linear fits to the differential abundance values.

\setcounter{table}{1}
\begin{table}
\centering
\caption{Differential parameters $\Delta$T$_{eff}$, $\Delta$log $g$, $\Delta$[Fe/H] and $\Delta$v$_{turb}$
derived between the star and their corresponding reference.}
%\hskip -0.35in
\scriptsize
\begin{tabular}{ccccc}
\hline
\hline
 (Star - Reference) & $\Delta$T$_{eff}$ & $\Delta$log $g$ & $\Delta$(Fe/H) & $\Delta$v$_{turb}$ \\
  & [K] & [dex] & [dex] & [km/s] \\
\hline
($\zeta^{1}$ Ret - Sun)              &  -67$\pm$29 & +0.09$\pm$0.05 & -0.195$\pm$0.005 & -0.11$\pm$0.27 \\
($\zeta^{2}$ Ret - Sun)              &  +77$\pm$28 & +0.10$\pm$0.04 & -0.215$\pm$0.004 & +0.04$\pm$0.09 \\
($\zeta^{1}$ Ret - $\zeta^{2}$ Ret)  & -144$\pm$22 & -0.01$\pm$0.03 & +0.020$\pm$0.003 & -0.15$\pm$0.07 \\
\hline
\end{tabular}
\normalsize
\label{param.dif}
\end{table}

\begin{figure}
\centering
\includegraphics[width=8cm]{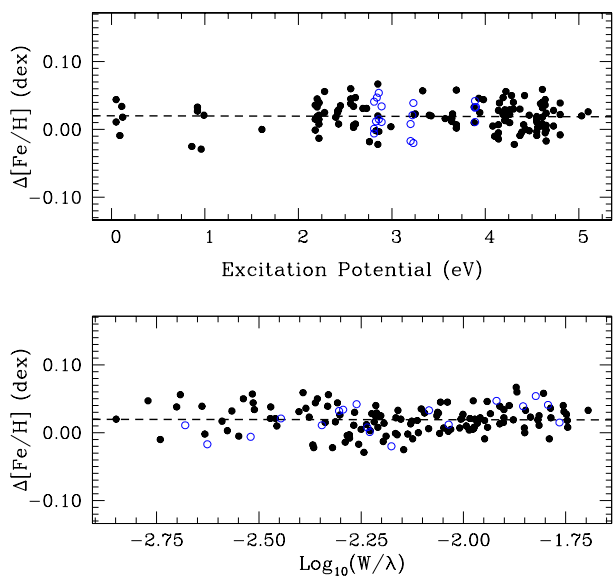}
\caption{Differential abundance vs excitation potential (upper panel) 
and differential abundance vs reduced EW (lower panel), for $\zeta^{1}$ Ret relative to $\zeta^{2}$ Ret.
Filled and empty points correspond to Fe I and Fe II, respectively.
The dashed line is a linear fit to the data.}
\label{equil-relat}%
\end{figure}

Then, the parameters of $\zeta^{1}$ Ret were recalculated but using $\zeta^{2}$ Ret as reference instead
of the Sun, i.e. ($\zeta^{1}$ Ret - $\zeta^{2}$ Ret).
Figure \ref{equil-relat} shows abundance vs excitation
potential and abundance vs EW$_{r}$, using similar symbols to those used in Figures
\ref{equil-zet01-sun} and \ref{equil-zet02-sun}.
The resulting stellar parameters for $\zeta^{1}$ Ret are the same to that obtained 
when we used the Sun as a reference, but with lower dispersions: {T$_{eff}$ = 5710$\pm$22 K},
{log $g$ = 4.53$\pm$0.03 dex}, {[Fe/H] = -0.195$\pm$0.003 dex,} and {v$_{turb}$ = 0.80$\pm$0.07 km/s}.
The differential parameters $\Delta$T$_{eff}$, $\Delta$log $g$, $\Delta$(Fe/H) and $\Delta$v$_{turb}$ derived between
$\zeta^{1}$ Ret and $\zeta^{2}$ Ret are also presented in the Table \ref{param.dif}.
Again, we found that the metallicity of $\zeta^{1}$ Ret is slightly higher than $\zeta^{2}$ Ret
by $\sim$ 0.02 dex.

Once the stellar parameters of the binary components were determined using iron lines,
we computed abundances for all remaining chemical elements. The hyperfine structure
splitting (HFS) was considered for V I, Mn I, Co I, Cu I, and Ba II, using the HFS
constants of \citet{kurucz-bell95} and performing spectral synthesis for these species.
We derived the O I abundances by using spectral synthesis with the line
6300.304 {\AA}\footnote{This line is blended with Ni I 6300.336 {\AA}.}.
(the O I triplet is out of the HARPS wavelength range), which is basically free of NLTE effects \citep{takeda03}.
Direct interpolation in the tables of \citet{amarsi15} results also in a null NLTE correction
for this O I line. We also applied NLTE corrections to Ba II (-0.10 dex) and Cu I (+0.04 dex)
in the same amount for both stars, interpolating in data of \citet{korotin11} and \citet{yan15}.

In Table \ref{table.abunds} we present the final differential abundances
[X/Fe]\footnote{We used the standard notation [X/Fe] $=$ [X/H] $-$ [Fe/H]}
of $\zeta^{1}$ Ret and $\zeta^{2}$ Ret relative to the Sun, and the differential abundances
of $\zeta^{1}$ Ret using $\zeta^{2}$ Ret as the reference star.
We present both the observational errors $\sigma_{obs}$
(estimated as $\sigma/\sqrt{(n-1)}$ , where $\sigma$ is the standard deviation of
the different lines) 
and systematic errors due to uncertainties in the stellar parameters
$\sigma_{par}$ (by adding quadratically the abundance variation when modifying
the stellar parameters by their uncertainties), as well as the total error
$\sigma_{TOT}$ obtained by quadratically adding  $\sigma_{obs}$,  $\sigma_{par}$ and
the error in [Fe/H].

\section{Results and discussion}

\begin{figure}
\centering
\includegraphics[width=8cm]{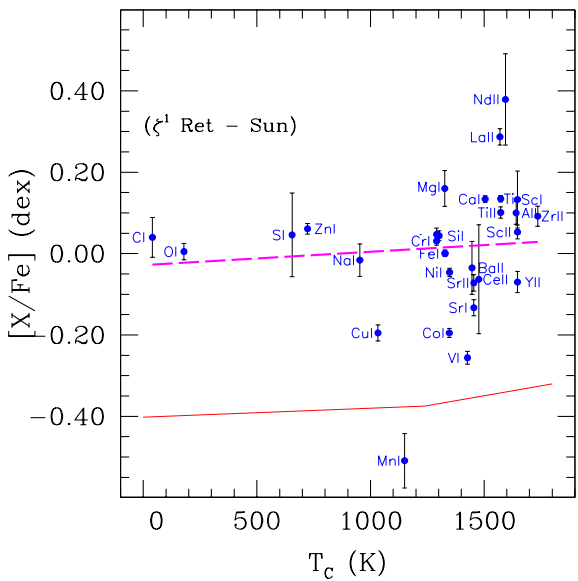}
\caption{Differential abundances {($\zeta^{1}$ Ret - Sun)} vs condensation temperature T$_{c}$.
The dashed line is a linear fit to the differential abundance values,
while the continuous line shows the solar-twins trend of \citet{melendez09}.}
\label{abund-zet01-sun}%
\end{figure}

\begin{figure}
\centering
\includegraphics[width=8cm]{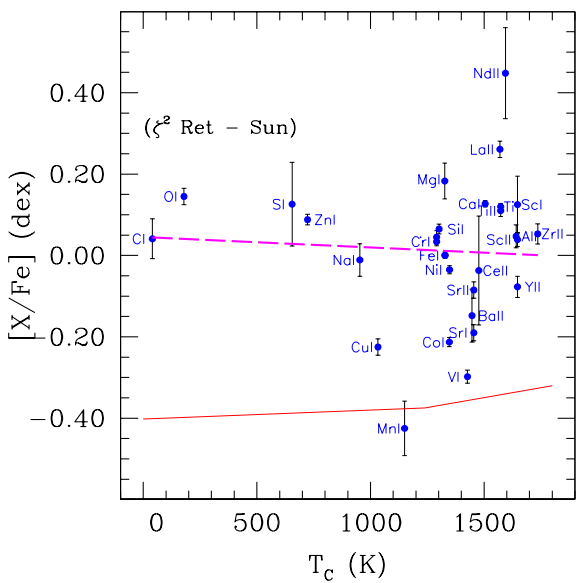}
\caption{Differential abundances {($\zeta^{2}$ Ret - Sun)} vs condensation temperature T$_{c}$.
The dashed line is a linear fit to the differential abundance values,
while the continuous line shows the solar-twins trend of \citet{melendez09}.}
\label{abund-zet02-sun}%
\end{figure}

The differential abundances of $\zeta^{1}$ Ret and $\zeta^{2}$ Ret relative to the Sun are presented
in Figures \ref{abund-zet01-sun} and \ref{abund-zet02-sun}, with condensation temperatures
taken from the 50\% T$_{c}$ values derived by \citet{lodders03}. 
We corrected by GCE effects when comparing (star-Sun) following the same procedure of
\citet{saffe15} by adopting the GCE fitting trends of \citet{gonz-hern13}. However, this
correction is discarded when comparing mutually the components of the binary system
($\zeta^{1}$ Ret - $\zeta^{2}$ Ret), due to their common natal environment.
Filled points in Figs. \ref{abund-zet01-sun} and \ref{abund-zet02-sun} correspond to the
differential abundances for the stars {$\zeta^{1}$ Ret} and {$\zeta^{2}$ Ret} relative to the Sun.
The continuous line correspond to the solar-twins trend of M09 (vertically shifted for comparison),
while the dashed line in Figs. \ref{abund-zet01-sun} and \ref{abund-zet02-sun} is a linear fit
to the abundance values.
We also present in the Table \ref{slopes} the derived slopes with their dispersions.
In order to provide another estimation of the significance of the slopes, we performed 100000
series of simulated random abundances and errors, following a similar method to MA15.
Then, assuming that the distribution of the simulated slopes follows a Gaussian distribution,
we compute the probability of the original slope "being by chance". This value is also
presented in the last column of Table \ref{slopes}.

\setcounter{table}{3}
\begin{table}
\centering
\caption{Derived slopes (abundance vs temperature condensation T$_{c}$), their dispersion
and probability of the slope "being by chance" (see text for details).}
%\hskip -0.35in
%\scriptsize
\begin{tabular}{rrr}
\hline
\hline
(Star - Reference)  & Slope$\pm \sigma$ & Prob. \\
  & [10$^{-5}$ dex/K] & \\
\hline
($\zeta^{1}$ Ret - Sun)                      & +3.31$\pm$2.23 & 0.48 \\
($\zeta^{2}$ Ret - Sun)                      & -2.57$\pm$1.99 & 0.52 \\
($\zeta^{1}$ Ret - $\zeta^{2}$ Ret)          & +6.27$\pm$1.55 & 0.27 \\
($\zeta^{1}$ Ret - $\zeta^{2}$ Ret)$_{Refr}$ & +3.85$\pm$1.02 & 0.29 \\
\hline
\end{tabular}
\normalsize
\label{slopes}
\end{table}

%We note that the general trends of $\zeta^{1}$ Ret and $\zeta^{2}$ Ret present slightly higher and lower slopes than a sun-like horizontal trend.
We note that the general trend of $\zeta^{1}$ Ret presents a slightly higher slope than the Sun,
while $\zeta^{2}$ Ret presents a slightly lower slope than the Sun.
This would correspond, for instance, to an slight lack of refractories (T$_{c}>$ 900 K) respect to
volatiles (T$_{c}<$ 900 K) when comparing $\zeta^{2}$ Ret with the Sun.
However, the general trends should be taken with caution due to the relatively high
dispersion of points (most elements spread between -0.20 dex and +0.20 dex).
These dispersions are greatly diminished by comparing mutually $\zeta^{1}$ Ret and $\zeta^{2}$ Ret,
which present the advantage of the strong similarity between them, together with
the independence of GCE and evolutionary effects.
Also, as we see in Table \ref{slopes}, the slopes of the stars relative to the Sun are derived
within $\sim$1.5$\sigma$, while for the case ($\zeta^{1}$ Ret - $\zeta^{2}$ Ret) the slope values
are within $\sim$4$\sigma$ i.e. the significance of the slope increases significatively.
The Table \ref{slopes} also includes the case of considering only the refractory elements between
$\zeta^{1}$ Ret and $\zeta^{2}$ Ret, showed as ($\zeta^{1}$ Ret - $\zeta^{2}$ Ret)$_{Refr}$.
The probability of the slopes "being by chance" are relatively high when the Sun is used
as reference ($\sim$\%50), however these values are diminished ($\sim$\%28) in the mutual
comparison ($\zeta^{1}$ Ret - $\zeta^{2}$ Ret), and are similar to those derived by MA15
(see e.g their Table 9).

The abundances of Mn I and Nd II seem to deviate from the general trends 
for both stars (see Figs. \ref{abund-zet01-sun} and \ref{abund-zet02-sun}).
We derived the abundances of Mn I including the hyperfine structure
splitting (HFS) using the constants of \citet{kurucz-bell95} in the spectral synthesis.
We do not detect an HFS abundance difference between the Mn I line 4502.21 {\AA} and other
Mn I lines, as reported by MA15.
For the Sun, \citet{bergemann-gehren07} estimated maximum NLTE corrections of +0.1 dex for
Mn, however even with such maximum correction the Mn abundance still results relatively low.
Given the similar parameters of our stars with the Sun, we consider that the NLTE effects
could also play a role in $\zeta^{1}$ Ret and $\zeta^{2}$ Ret.
The abundance of Nd II was derived using the Equivalent Widths of two lines
(4021.33 {\AA} and 4446.38 {\AA}). We have no evidence of an obvious blend at these lines,
however it is difficult to discard this possibility. Other Nd II lines such as 4059.95 {\AA}
or 5234.19 {\AA} are very weak. \citet{mashonkina05} studied NLTE effects of Nd II but in
stars with higher temperatures ($>$7500 K).
Then, in order to derive representative trends we excluded Mn I and Nd II from the
linear fits.

\begin{figure}
\centering
\includegraphics[width=8cm]{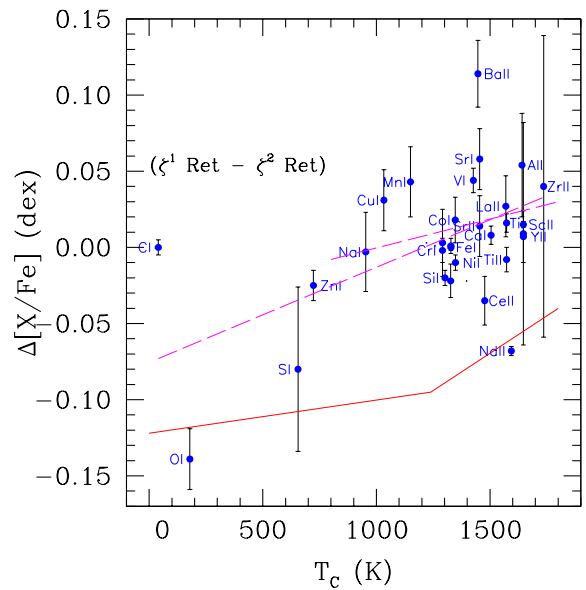}
\caption{Differential abundances ($\zeta^{1}$ Ret - $\zeta^{2}$ Ret) vs condensation temperature T$_{c}$.
The long-dashed lines are linear fits to all species and to the refractory species.
The solar-twins trend of \citet{melendez09} is shown with a continuous line.}
\label{relat.tc}%
\end{figure}

The differential abundances of $\zeta^{1}$ Ret using $\zeta^{2}$ Ret as reference are
presented in Fig. \ref{relat.tc}. In our calculation, this plot corresponds
to the abundance values derived with the highest possible precision.
The continuous line in this Figure presents the solar-twins trend of M09 (vertically shifted),
while the long-dashed lines are linear fits to all elements as well as the refractory elements.
The differential abundances of the refractories (average of +0.016 dex)
are higher than the volatiles (average of -0.061 dex), with a general slope of
{6.27$\pm$1.55 10$^{-5}$ dex/K}, as we see in the Fig. \ref{relat.tc}.
Although the general trend seems to be driven by O I (which present a relatively low 
abundance value), when excluding O I the slope results {3.96$\pm$1.05 10$^{-5}$ dex/K} i.e. a general
trend even closer to the refractory trend.
The refractory elements (alone) also show a trend with T$_{c}$, with a slope of
{3.85$\pm$1.02 10$^{-5}$ dex/K}.
For comparison, the slope of refractories between the components of the binary system
{16 Cyg} resulted {1.88 10$^{-5}$ dex/K} and clearly showing a higher abundance in
refractory than volatile elements \citep{tucci-maia14}.
We caution, however, that there is still no total consensus on the possible chemical
differences between the components of 16 Cyg
\citep[e.g.][]{takeda05,schuler11b,tucci-maia14}.
Then, the higher value of the refractory elements compared to volatile elements, together
with the positive slope in the trend of refractory elements with T$_{c}$, point toward
a lack of refractory elements in $\zeta^{2}$ Ret relative to $\zeta^{1}$ Ret.
% similar to the case of the binary system 16 Cyg \citep{tucci-maia14}.

Detailed abundance studies performed on binary systems with similar components
(where at least one component hosts a planet) showed different results.
Some binary systems such as HAT-P-1 and HD 80606 does not seem to present significant
T$_{c}$ trends between their components \citep{liu14,saffe15}.
On the other hand, the binary systems XO-2 and HD 20781 seem to present a relative
T$_{c}$ trend \citep{biazzo15,mack14}, although they are particular cases, given that
both components of the system host a planet. 
Also, possible metallicity differences of wide binary stars and multiple systems
have been studied in the literature. Most of the cases present almost no metallicity
differences, similar to the case of the triple system HD 132563 \citep{desidera11}.
However, $\sim$17\% of the wide binaries do present slight metallicity differences 
between their components \citep{desidera04,desidera06}.
The origin of these slight differences is not totally clear, and a possible explanation
lies the planet formation process \citep[e.g. ][]{desidera04,desidera06}.

Following the interpretation of M09 and R10, the lack of refractory elements in $\zeta^{2}$ Ret
compared to $\zeta^{1}$ Ret could be identified with the signature of the planet formation process.
Because no planets are detected around the stars of this binary system, the refractory
elements depleted in $\zeta^{2}$ Ret are possibly locked-up in rocky bodies (e.g. planetesimals
and/or asteroids) whose colisions could produce the bright debris disk observed in this object.
In fact, the slightly lower metallicity of $\zeta^{2}$ Ret compared to their companion
(by $\sim$ 0.02 dex) is also compatible with this scenario.
Probably, the relatively low metallicity of both stars prevented the formation of giant planets
in this binary system. However the presence of circumstellar material around $\zeta^{2}$ Ret
(the debris disk) has been confirmed, as we mentioned previously.
The fact that the conditions required to form a debris-disk are more easily met than the
conditions to form gas-giant planets, is in agreement with the core-accretion model of planet formation
\citep[e.g.][]{pollack96,mordasini12}.

The rocky mass of depleted material in $\zeta^{2}$ Ret was estimated following \citet{chambers10},
in order to reproduce the trend of the refractory elements of Fig. \ref{relat.tc}.
Using a convection zone similar to the Sun (M$_{cz}$ = 0.023 M$_{\sun}$) we obtain
M$_{rock} \sim$ 3 M$_{\oplus}$.
However, at the time of the planet formation process M$_{cz}$ could be greater than this value.
For instance, adopting M$_{cz}$ = 0.050 M$_{\sun}$ we derive M$_{rock} \sim$ 7 M$_{\oplus}$.
Then, M$_{rock}\sim$ 3 M$_{\oplus}$ should be considered as a lower limit.
On the oher hand, there is no direct estimation for the debris disk mass of $\zeta^{2}$ Ret
\citep[see e.g.][]{eiroa10,faramaz14}. The most precise estimates debris disk masses comes
from sub-mm observations \citep[see e.g.][]{wyatt08}.
\citet{krivov08} modeled 5 solar-type debris-disk stars and fitted the far-IR emission 
using disk masses in the range 3-50 M$_{\oplus}$ and radii between 100-200 AU.
The mass range and radii are compatible with the mass derived from \citet{chambers10}
and the observed location of the disk around $\zeta^{2}$ Ret \citep[R $\sim$ 100 AU,][]{eiroa10}.
However, \citet{krivov08} caution that the mass and location of the debris disk depend
significatively on the colisional model and grain properties adopted. 

\section{Conclusions}

We performed a high-precision differential abundance determination in both components
of the remarkable binary system {$\zeta^{1}$ Ret - $\zeta^{2}$ Ret}, in order to detect a
possible T$_{c}$ trend.
Both stars present very similar stellar parameters, which greatly diminishes
the errors in the abundance determination, GCE or evolutionary effects.
The star $\zeta^{2}$ Ret hosts a debris disk, while there is no debris disk detected 
(nor a planet) around $\zeta^{1}$ Ret.
First, we derived stellar parameters and differential abundances of both stars using the
Sun as the reference star and then for $\zeta^{1}$ Ret using $\zeta^{2}$ Ret as reference.
Our calculation included NLTE corrections for Ba II and Cu I as well as GCE
corrections for all chemical species, where the Sun was used as reference.
We compared the possible temperature condensation T$_{c}$ trends of the stars
with the solar-twins trend of \citet{melendez09}.

In comparing the stars to each other, $\zeta^{1}$ Ret resulted slightly more metal rich
than $\zeta^{2}$ Ret by $\sim$ 0.02 dex. Also, the differential abundances of the
refractories resulted higher than the volatiles, and the general trend of the refractory
elements with T$_{c}$ showed a positive slope.
These facts together point toward a lack of refractory elements in $\zeta^{2}$ Ret relative
to $\zeta^{1}$ Ret, similar to the case of the binary system 16 Cyg \citep{tucci-maia14}.
We caution, however, that there is still no total consensus on the possible chemical
differences between the components of 16 Cyg
\citep[e.g.][]{takeda05,schuler11b,tucci-maia14}.
We note that the statistical result of \citet{maldonado15} does not exclude possible
Tc trends in particular stars such as the $\zeta$ Ret system.
Then, following the interpretation of M09 and R10, we propose an scenario in which the refractory
elements depleted in $\zeta^{2}$ Ret are possibly locked-up in the rocky material that orbits
this star and produce the debris disk observed around this object.
We estimated a lower limit of M$_{rock} \sim$ 3 M$_{\oplus}$ for the rocky mass of
depleted material according to \citet{chambers10}, which is compatible with a rough
estimation of 3-50 M$_{\oplus}$ of a debris disk mass around a solar-type star \citep{krivov08}.
We strongly encourage high-precision abundance studies in binary systems
with similar components, which is a crucial tool for helping to detect the
possible chemical pattern of the planet formation process.

\begin{acknowledgements}
The authors thank Drs. R. Kurucz and C. Sneden for making their codes available to us.
We also thank the anonymous referee for constructive comments that improved the paper.
\end{acknowledgements}

\onltab{1}{
\onecolumn
\setcounter{table}{0}
\begin{longtable}{ccccccc}
\caption{Line list used in this work. The columns present the element, 
wavelength $\lambda$, excitation potential (EP), log $gf$, equivalent widths
of $\zeta^{1}$ Ret, $\zeta^{2}$ Ret, and Sun ($EW_{1}$, $EW_{2}$ , and $EW_{Sun}$).
The abundances of lines without EWs were measured using synthetic spectra.}
\\
\hline \hline
Element  & $\lambda$  & EP   & log $gf$ & $EW_{1}$ & $EW_{2}$ & $EW_{Sun}$ \\
         & [\AA]      & [eV] & [dex]  & [m\AA]   & [m\AA]   & [m\AA]     \\
\hline
\endfirsthead
\caption{Continued.}\\
\hline \hline
Element  & $\lambda$  & EP   & log $gf$ & $EW_{1}$ & $EW_{2}$ & $EW_{Sun}$ \\
         & [\AA]      & [eV] & [dex]  & [m\AA]   & [m\AA]   & [m\AA]     \\
\hline
\endhead
\hline
\endfoot
%datos tomados de line.by.line/progs/salida.final.txt
    6.00 &   5052.167 &    7.680 &   -1.240 &    23.3 &    25.4 &    32.9 \\
    6.00 &   6587.610 &    8.540 &   -1.050 &     8.3 &     9.6 &    11.3 \\
    8.00 &   6300.304 &    0.000 &   -9.820 \\
   11.00 &   4751.822 &    2.100 &   -2.080 &     9.4 &     7.7 &    12.9 \\
   11.00 &   5148.838 &    2.100 &   -2.040 &     9.0 &     6.7 &    11.5 \\
   11.00 &   6154.225 &    2.100 &   -1.550 &    24.3 &    23.0 &    38.2 \\
   11.00 &   6160.747 &    2.100 &   -1.250 &    42.8 &    36.9 &    53.8 \\
   12.00 &   4571.095 &    0.000 &   -5.620 &   102.8 &    97.2 &   107.0 \\
   12.00 &   4730.040 &    4.340 &   -2.390 &    58.2 &    54.0 &    67.1 \\
   12.00 &   5711.088 &    4.340 &   -1.730 &   102.2 &    94.8 &   103.5 \\
   12.00 &   6318.717 &    5.110 &   -1.950 &    32.7 &    28.7 &    38.2 \\
   12.00 &   6319.236 &    5.110 &   -2.160 &    29.8 &    26.3 &    25.9 \\
   13.00 &   5557.070 &    3.140 &   -2.210 &     9.3 &     6.1 &    12.7 \\
   13.00 &   6696.018 &    3.140 &   -1.480 &    30.9 &    25.9 &    34.8 \\
   13.00 &   6698.667 &    3.140 &   -1.780 &    15.7 &    12.1 &    20.7 \\
   14.00 &   5488.983 &    5.610 &   -1.690 &    11.0 &    11.9 &    17.2 \\
   14.00 &   5517.540 &    5.080 &   -2.500 &     9.6 &     8.1 &    13.7 \\
   14.00 &   5645.611 &    4.930 &   -2.040 &    27.3 &    25.8 &    36.1 \\
   14.00 &   5665.554 &    4.920 &   -1.940 &    31.9 &    28.9 &    40.9 \\
   14.00 &   5684.484 &    4.950 &   -1.550 &    49.7 &    46.8 &    60.6 \\
   14.00 &   5690.425 &    4.930 &   -1.770 &    38.5 &    36.5 &    50.5 \\
   14.00 &   5701.104 &    4.930 &   -1.950 &    30.4 &    28.7 &    37.5 \\
   14.00 &   5753.640 &    5.620 &   -1.330 &    36.0 &    34.8 &    42.3 \\
   14.00 &   5772.145 &    5.082 &   -1.653 &    42.2 &    41.7 &    45.8 \\
   14.00 &   5793.073 &    4.930 &   -1.960 &    35.0 &    31.8 &    42.2 \\
   14.00 &   5948.540 &    5.080 &   -1.208 &    80.0 &    77.8 &    86.0 \\
   14.00 &   6125.021 &    5.610 &   -1.500 &    23.1 &    21.5 &    29.7 \\
   14.00 &   6142.490 &    5.620 &   -1.540 &    26.9 &    25.7 &    31.0 \\
   14.00 &   6145.015 &    5.620 &   -1.410 &    27.4 &    25.9 &    37.3 \\
   14.00 &   6195.460 &    5.870 &   -1.666 &    10.1 &     9.6 &    15.7 \\
   14.00 &   6237.330 &    5.610 &   -1.116 &    50.4 &    47.0 &    59.2 \\
   14.00 &   6243.823 &    5.620 &   -1.270 &    34.7 &    33.5 &    41.9 \\
   14.00 &   6244.476 &    5.620 &   -1.320 &    34.4 &    33.9 &    50.6 \\
   14.00 &   6527.210 &    5.870 &   -1.230 &    33.6 &    32.0 &    40.8 \\
   14.00 &   6721.848 &    5.860 &   -1.120 &    30.8 &    30.7 &    44.8 \\
   14.00 &   6741.630 &    5.980 &   -1.650 &     9.3 &     9.4 &    17.1 \\
   16.00 &   4695.443 &    6.530 &   -1.830 &     4.9 &     5.7 &     5.6 \\
   16.00 &   6052.656 &    7.870 &   -0.400 &     4.7 &     6.8 &     7.4 \\
   20.00 &   4512.268 &    2.530 &   -1.900 &    18.4 &    15.6 &    21.6 \\
   20.00 &   5260.387 &    2.520 &   -1.720 &    27.2 &    23.0 &    31.6 \\
   20.00 &   5261.710 &    2.520 &   -0.680 &    95.6 &    89.6 &   100.0 \\
   20.00 &   5512.980 &    2.930 &   -0.460 &    81.4 &    73.4 &    83.0 \\
   20.00 &   5581.965 &    2.520 &   -0.560 &    91.6 &    84.7 &    92.9 \\
   20.00 &   5590.114 &    2.520 &   -0.570 &    88.5 &    79.7 &    89.8 \\
   20.00 &   5867.562 &    2.930 &   -1.570 &    18.5 &    13.7 &    24.2 \\
   20.00 &   6156.020 &    2.520 &   -2.497 &     6.4 &     5.1 &     8.4 \\
   20.00 &   6161.297 &    2.520 &   -1.270 &    55.1 &    49.2 &    58.2 \\
   20.00 &   6166.439 &    2.520 &   -1.140 &    62.0 &    55.5 &    69.5 \\
   20.00 &   6169.042 &    2.520 &   -0.800 &    88.4 &    79.7 &    92.5 \\
   20.00 &   6169.550 &    2.520 &   -0.580 &   110.3 &    98.7 &   110.4 \\
   20.00 &   6455.598 &    2.520 &   -1.340 &    50.6 &    43.0 &    56.7 \\
   20.00 &   6471.662 &    2.530 &   -0.690 &    85.7 &    80.3 &    88.9 \\
   20.00 &   6499.650 &    2.520 &   -0.820 &    80.3 &    73.8 &    85.1 \\
   20.00 &   6572.800 &    0.000 &   -4.280 &    27.4 &    19.9 &    33.2 \\
   21.00 &   4743.821 &    1.450 &    0.350 &     7.5 &     4.8 &     8.1 \\
   21.00 &   5671.821 &    1.450 &    0.550 &    10.7 &     8.6 &    13.9 \\
   21.10 &   5526.820 &    1.770 &    0.140 &    68.8 &    69.0 &    76.2 \\
   21.10 &   5657.870 &    1.510 &   -0.300 &    59.8 &    59.8 &    65.7 \\
   21.10 &   5667.140 &    1.500 &   -1.020 &    25.6 &    24.5 &    34.6 \\
   21.10 &   6245.630 &    1.510 &   -1.030 &    28.2 &    26.8 &    36.5 \\
   21.10 &   6604.578 &    1.360 &   -1.150 &    28.3 &    26.9 &    37.1 \\
   22.00 &   4465.802 &    1.740 &   -0.160 &    35.0 &    27.2 &    38.6 \\
   22.00 &   4512.733 &    0.840 &   -0.420 &    64.3 &    55.7 &    66.0 \\
   22.00 &   4555.485 &    0.850 &   -0.490 &    59.8 &    54.4 &    63.6 \\
   22.00 &   4617.280 &    1.750 &    0.450 &    60.9 &    53.4 &    62.4 \\
   22.00 &   4623.100 &    1.740 &    0.170 &    53.8 &    46.0 &    60.9 \\
   22.00 &   4645.190 &    1.730 &   -0.670 &    20.4 &    15.2 &    20.9 \\
   22.00 &   4656.470 &    0.000 &   -1.310 &    67.7 &    58.8 &    68.7 \\
   22.00 &   4722.610 &    1.050 &   -1.430 &    16.1 &    12.3 &    15.9 \\
   22.00 &   4758.120 &    2.250 &    0.430 &    39.4 &    33.0 &    43.0 \\
   22.00 &   4759.272 &    2.260 &    0.510 &    42.3 &    36.5 &    47.0 \\
   22.00 &   4778.258 &    2.240 &   -0.220 &    12.0 &     9.7 &    16.7 \\
   22.00 &   4820.410 &    1.500 &   -0.440 &    39.5 &    31.5 &    43.8 \\
   22.00 &   4840.880 &    0.900 &   -0.450 &    61.7 &    57.3 &    65.8 \\
   22.00 &   4913.616 &    1.870 &    0.160 &    48.5 &    39.9 &    49.7 \\
   22.00 &   4999.500 &    0.830 &    0.270 &   101.9 &    93.3 &   103.3 \\
   22.00 &   5022.871 &    0.830 &   -0.430 &    70.1 &    62.9 &    72.9 \\
   22.00 &   5024.850 &    0.820 &   -0.560 &    64.4 &    57.0 &    70.2 \\
   22.00 &   5039.960 &    0.020 &   -1.200 &    76.0 &    69.5 &    75.7 \\
   22.00 &   5071.490 &    1.460 &   -0.800 &    24.8 &    20.8 &    27.3 \\
   22.00 &   5113.448 &    1.440 &   -0.780 &    24.6 &    19.1 &    27.1 \\
   22.00 &   5147.479 &    0.000 &   -2.010 &    33.0 &    24.7 &    35.7 \\
   22.00 &   5219.700 &    0.020 &   -2.240 &    24.1 &    17.1 &    28.3 \\
   22.00 &   5295.774 &    1.070 &   -1.630 &    10.8 &     8.0 &    13.4 \\
   22.00 &   5471.200 &    1.440 &   -1.400 &     7.2 &     5.2 &     7.4 \\
   22.00 &   5490.150 &    1.460 &   -0.930 &    18.1 &    14.4 &    20.6 \\
   22.00 &   5689.459 &    2.300 &   -0.360 &     9.6 &     6.9 &    12.9 \\
   22.00 &   5739.464 &    2.250 &   -0.600 &     6.3 &     4.6 &     7.7 \\
   22.00 &   5866.452 &    1.070 &   -0.840 &    42.7 &    35.1 &    48.9 \\
   22.00 &   5965.840 &    1.880 &   -0.490 &    27.4 &    20.4 &    27.2 \\
   22.00 &   5978.550 &    1.870 &   -0.602 &    20.0 &    14.6 &    22.2 \\
   22.00 &   6064.630 &    1.050 &   -1.959 &     6.8 &     4.4 &     7.9 \\
   22.00 &   6091.174 &    2.270 &   -0.420 &    12.2 &     8.2 &    15.2 \\
   22.00 &   6126.217 &    1.070 &   -1.420 &    18.7 &    13.4 &    20.8 \\
   22.00 &   6258.104 &    1.440 &   -0.350 &    47.2 &    38.3 &    50.9 \\
   22.00 &   6261.101 &    1.430 &   -0.480 &    42.4 &    35.2 &    49.1 \\
   22.00 &   6312.234 &    1.460 &   -1.496 &     5.7 &     3.1 &     8.0 \\
   22.00 &   6599.104 &    0.900 &   -2.029 &     7.1 &     5.3 &     8.9 \\
   22.00 &   6743.130 &    0.899 &   -1.630 &    15.1 &    10.6 &    18.6 \\
   22.10 &   4470.857 &    1.160 &   -2.060 &    61.0 &    62.3 &    65.0 \\
   22.10 &   4544.028 &    1.240 &   -2.530 &    36.4 &    36.9 &    42.1 \\
   22.10 &   4609.265 &    1.180 &   -3.430 &    11.4 &    11.8 &    17.1 \\
   22.10 &   4657.212 &    1.240 &   -2.470 &    44.5 &    45.5 &    48.9 \\
   22.10 &   4779.985 &    2.048 &   -1.260 &    60.2 &    60.2 &    64.8 \\
   22.10 &   4798.532 &    1.080 &   -2.670 &    38.7 &    38.8 &    45.0 \\
   22.10 &   4865.611 &    1.120 &   -2.810 &    33.4 &    33.1 &    41.5 \\
   22.10 &   4874.014 &    3.100 &   -0.900 &    30.9 &    32.1 &    36.7 \\
   22.10 &   4911.193 &    3.120 &   -0.540 &    47.3 &    45.9 &    51.1 \\
   22.10 &   5381.015 &    1.570 &   -1.970 &    50.8 &    51.5 &    59.6 \\
   22.10 &   5418.767 &    1.580 &   -2.110 &    42.7 &    42.5 &    48.8 \\
   22.10 &   5490.690 &    1.570 &   -2.430 &    17.4 &    17.2 &    21.0 \\
   23.00 &   4875.442 &    0.040 &   -3.375 \\
   23.00 &   4875.454 &    0.040 &   -2.260 \\
   23.00 &   4875.461 &    0.040 &   -2.964 \\
   23.00 &   4875.468 &    0.040 &   -1.420 \\
   23.00 &   4875.471 &    0.040 &   -2.064 \\
   23.00 &   4875.477 &    0.040 &   -2.742 \\
   23.00 &   4875.483 &    0.040 &   -1.561 \\
   23.00 &   4875.485 &    0.040 &   -2.010 \\
   23.00 &   4875.491 &    0.040 &   -2.617 \\
   23.00 &   4875.495 &    0.040 &   -1.725 \\
   23.00 &   4875.497 &    0.040 &   -2.032 \\
   23.00 &   4875.502 &    0.040 &   -2.566 \\
   23.00 &   4875.505 &    0.040 &   -1.923 \\
   23.00 &   4875.506 &    0.040 &   -2.123 \\
   23.00 &   4875.509 &    0.040 &   -2.596 \\
   23.00 &   4875.511 &    0.040 &   -2.178 \\
   23.00 &   4875.511 &    0.040 &   -2.311 \\
   23.00 &   4875.515 &    0.040 &   -2.566 \\
   23.00 &   5703.555 &    1.050 &   -0.777 \\
   23.00 &   5703.569 &    1.050 &   -0.993 \\
   23.00 &   5703.569 &    1.050 &   -1.403 \\
   23.00 &   5703.580 &    1.050 &   -1.242 \\
   23.00 &   5703.580 &    1.050 &   -1.276 \\
   23.00 &   5703.581 &    1.050 &   -2.268 \\
   23.00 &   5703.589 &    1.050 &   -1.250 \\
   23.00 &   5703.589 &    1.050 &   -1.715 \\
   23.00 &   5703.590 &    1.050 &   -1.840 \\
   23.00 &   5703.596 &    1.050 &   -1.414 \\
   23.00 &   5703.596 &    1.050 &   -1.590 \\
   23.00 &   5703.601 &    1.050 &   -1.414 \\
   23.00 &   5727.008 &    1.080 &   -0.693 \\
   23.00 &   5727.016 &    1.080 &   -1.701 \\
   23.00 &   5727.022 &    1.080 &   -3.003 \\
   23.00 &   5727.028 &    1.080 &   -0.798 \\
   23.00 &   5727.035 &    1.080 &   -1.490 \\
   23.00 &   5727.040 &    1.080 &   -2.605 \\
   23.00 &   5727.045 &    1.080 &   -0.914 \\
   23.00 &   5727.051 &    1.080 &   -1.417 \\
   23.00 &   5727.056 &    1.080 &   -2.400 \\
   23.00 &   5727.060 &    1.080 &   -1.043 \\
   23.00 &   5727.065 &    1.080 &   -1.411 \\
   23.00 &   5727.069 &    1.080 &   -2.303 \\
   23.00 &   5727.072 &    1.080 &   -1.189 \\
   23.00 &   5727.075 &    1.080 &   -1.458 \\
   23.00 &   5727.078 &    1.080 &   -2.303 \\
   23.00 &   5727.081 &    1.080 &   -1.359 \\
   23.00 &   5727.084 &    1.080 &   -1.563 \\
   23.00 &   5727.086 &    1.080 &   -2.458 \\
   23.00 &   5727.087 &    1.080 &   -1.563 \\
   23.00 &   5727.089 &    1.080 &   -1.759 \\
   23.00 &   5727.091 &    1.080 &   -1.826 \\
   23.00 &   5727.619 &    1.050 &   -1.456 \\
   23.00 &   5727.619 &    1.050 &   -1.867 \\
   23.00 &   5727.653 &    1.050 &   -1.753 \\
   23.00 &   5727.653 &    1.050 &   -2.072 \\
   23.00 &   5727.654 &    1.050 &   -1.867 \\
   23.00 &   5727.681 &    1.050 &   -1.753 \\
   23.00 &   5727.681 &    1.050 &   -1.878 \\
   23.00 &   5727.681 &    1.050 &   -9.850 \\
   23.00 &   5727.701 &    1.050 &   -2.054 \\
   23.00 &   5727.702 &    1.050 &   -1.878 \\
   23.00 &   6081.417 &    1.050 &   -1.814 \\
   23.00 &   6081.418 &    1.050 &   -1.638 \\
   23.00 &   6081.428 &    1.050 &   -1.638 \\
   23.00 &   6081.428 &    1.050 &   -9.610 \\
   23.00 &   6081.429 &    1.050 &   -1.513 \\
   23.00 &   6081.443 &    1.050 &   -1.513 \\
   23.00 &   6081.443 &    1.050 &   -1.832 \\
   23.00 &   6081.444 &    1.050 &   -1.627 \\
   23.00 &   6081.461 &    1.050 &   -1.627 \\
   23.00 &   6081.462 &    1.050 &   -1.216 \\
   23.00 &   6090.194 &    1.080 &   -0.700 \\
   23.00 &   6090.201 &    1.080 &   -0.841 \\
   23.00 &   6090.207 &    1.080 &   -1.005 \\
   23.00 &   6090.208 &    1.080 &   -1.540 \\
   23.00 &   6090.213 &    1.080 &   -1.203 \\
   23.00 &   6090.213 &    1.080 &   -1.344 \\
   23.00 &   6090.217 &    1.080 &   -1.290 \\
   23.00 &   6090.217 &    1.080 &   -1.458 \\
   23.00 &   6090.220 &    1.080 &   -2.655 \\
   23.00 &   6090.221 &    1.080 &   -1.312 \\
   23.00 &   6090.221 &    1.080 &   -1.846 \\
   23.00 &   6090.223 &    1.080 &   -1.403 \\
   23.00 &   6090.223 &    1.080 &   -2.244 \\
   23.00 &   6090.225 &    1.080 &   -1.591 \\
   23.00 &   6090.225 &    1.080 &   -2.022 \\
   23.00 &   6090.226 &    1.080 &   -1.897 \\
   23.00 &   6090.227 &    1.080 &   -1.846 \\
   23.00 &   6090.227 &    1.080 &   -1.876 \\
   23.00 &   6111.592 &    1.042 &   -1.701 \\
   23.00 &   6111.632 &    1.042 &   -1.224 \\
   23.00 &   6111.656 &    1.042 &   -1.224 \\
   23.00 &   6111.696 &    1.042 &   -1.370 \\
   23.00 &   6119.528 &    1.063 &   -0.360 \\
   23.00 &   6199.149 &    0.286 &   -2.133 \\
   23.00 &   6199.167 &    0.286 &   -2.238 \\
   23.00 &   6199.182 &    0.286 &   -2.354 \\
   23.00 &   6199.197 &    0.286 &   -2.483 \\
   23.00 &   6199.201 &    0.286 &   -3.141 \\
   23.00 &   6199.209 &    0.286 &   -2.629 \\
   23.00 &   6199.212 &    0.286 &   -2.930 \\
   23.00 &   6199.221 &    0.286 &   -2.799 \\
   23.00 &   6199.221 &    0.286 &   -2.857 \\
   23.00 &   6199.229 &    0.286 &   -2.851 \\
   23.00 &   6199.230 &    0.286 &   -3.003 \\
   23.00 &   6199.235 &    0.286 &   -2.898 \\
   23.00 &   6199.238 &    0.286 &   -3.266 \\
   23.00 &   6199.240 &    0.286 &   -3.003 \\
   23.00 &   6199.243 &    0.286 &   -3.199 \\
   23.00 &   6199.246 &    0.286 &   -4.443 \\
   23.00 &   6199.251 &    0.286 &   -4.045 \\
   23.00 &   6199.253 &    0.286 &   -3.840 \\
   23.00 &   6199.253 &    0.286 &   -3.898 \\
   23.00 &   6199.255 &    0.286 &   -3.743 \\
   23.00 &   6199.255 &    0.286 &   -3.743 \\
   23.00 &   6242.798 &    0.262 &   -2.054 \\
   23.00 &   6242.798 &    0.262 &   -2.521 \\
   23.00 &   6242.829 &    0.262 &   -2.375 \\
   23.00 &   6242.837 &    0.262 &   -2.375 \\
   23.00 &   6242.852 &    0.262 &   -2.396 \\
   23.00 &   6242.868 &    0.262 &   -2.852 \\
   23.00 &   6243.045 &    0.300 &   -2.712 \\
   23.00 &   6243.060 &    0.300 &   -2.497 \\
   23.00 &   6243.075 &    0.300 &   -2.420 \\
   23.00 &   6243.087 &    0.300 &   -1.649 \\
   23.00 &   6243.087 &    0.300 &   -2.409 \\
   23.00 &   6243.097 &    0.300 &   -1.785 \\
   23.00 &   6243.099 &    0.300 &   -2.452 \\
   23.00 &   6243.106 &    0.300 &   -1.933 \\
   23.00 &   6243.109 &    0.300 &   -2.555 \\
   23.00 &   6243.114 &    0.300 &   -2.092 \\
   23.00 &   6243.118 &    0.300 &   -2.776 \\
   23.00 &   6243.120 &    0.300 &   -2.261 \\
   23.00 &   6243.125 &    0.300 &   -2.428 \\
   23.00 &   6243.129 &    0.300 &   -2.566 \\
   23.00 &   6243.132 &    0.300 &   -2.580 \\
   23.00 &   6243.140 &    0.300 &   -2.712 \\
   23.00 &   6243.142 &    0.300 &   -2.776 \\
   23.00 &   6243.143 &    0.300 &   -2.497 \\
   23.00 &   6243.145 &    0.300 &   -2.555 \\
   23.00 &   6243.146 &    0.300 &   -2.420 \\
   23.00 &   6243.146 &    0.300 &   -2.452 \\
   23.00 &   6243.147 &    0.300 &   -2.409 \\
   23.00 &   6285.160 &    0.275 &   -1.540 \\
   24.00 &   4511.900 &    3.090 &   -0.390 &    33.7 &    28.8 &    40.9 \\
   24.00 &   4545.945 &    0.940 &   -1.310 &    81.5 &    74.5 &    84.7 \\
   24.00 &   4626.180 &    0.970 &   -1.470 &    79.8 &    74.1 &    82.7 \\
   24.00 &   4700.599 &    2.710 &   -1.250 &    11.2 &     9.8 &    14.0 \\
   24.00 &   4708.017 &    3.170 &    0.090 &    50.9 &    45.9 &    55.5 \\
   24.00 &   4767.860 &    3.560 &   -0.600 &    12.7 &    11.1 &    17.4 \\
   24.00 &   4775.140 &    3.550 &   -1.020 &     5.2 &     4.9 &     7.5 \\
   24.00 &   4789.340 &    2.540 &   -0.350 &    57.6 &    51.2 &    64.8 \\
   24.00 &   4801.047 &    3.120 &   -0.130 &    42.7 &    39.1 &    49.0 \\
   24.00 &   4936.335 &    3.110 &   -0.250 &    36.6 &    32.4 &    43.9 \\
   24.00 &   5214.140 &    3.370 &   -0.740 &    13.4 &    11.6 &    16.7 \\
   24.00 &   5238.964 &    2.710 &   -1.270 &    12.1 &    10.7 &    15.8 \\
   24.00 &   5247.566 &    0.960 &   -1.590 &    76.6 &    67.7 &    81.5 \\
   24.00 &   5272.007 &    3.450 &   -0.420 &    18.7 &    15.1 &    21.5 \\
   24.00 &   5287.200 &    3.440 &   -0.870 &     7.9 &     5.8 &    10.8 \\
   24.00 &   5296.691 &    0.980 &   -1.360 &    89.8 &    80.4 &    94.2 \\
   24.00 &   5300.744 &    0.980 &   -2.130 &    49.5 &    43.9 &    58.3 \\
   24.00 &   5345.801 &    1.000 &   -0.950 &   110.6 &    99.1 &   113.5 \\
   24.00 &   5348.312 &    1.000 &   -1.210 &    93.7 &    85.0 &    98.3 \\
   24.00 &   5628.621 &    3.420 &   -0.760 &    11.2 &     7.7 &    13.6 \\
   24.00 &   5783.080 &    3.320 &   -0.430 &    23.1 &    19.6 &    29.8 \\
   24.00 &   5783.870 &    3.320 &   -0.290 &    33.5 &    28.1 &    45.3 \\
   24.00 &   5787.930 &    3.322 &   -0.080 &    36.0 &    32.9 &    46.1 \\
   24.00 &   6330.100 &    0.941 &   -2.900 &    19.4 &    15.0 &    26.5 \\
   24.00 &   6661.080 &    4.190 &   -0.190 &     7.3 &     6.2 &    13.0 \\
   24.00 &   6882.477 &    3.438 &   -0.375 &    24.8 &    18.6 &    31.9 \\
   24.00 &   6882.997 &    3.438 &   -0.420 &    26.9 &    23.9 &    29.2 \\
   24.10 &   4588.199 &    4.070 &   -0.590 &    63.9 &    65.2 &    69.6 \\
   24.10 &   4592.049 &    4.070 &   -1.250 &    41.0 &    41.9 &    47.7 \\
   24.10 &   4616.628 &    4.070 &   -1.210 &    38.7 &    40.4 &    47.2 \\
   24.10 &   5237.328 &    4.070 &   -1.090 &    44.8 &    47.8 &    52.8 \\
   24.10 &   5246.767 &    3.710 &   -2.440 &    11.6 &    10.2 &    14.0 \\
   24.10 &   5502.067 &    4.170 &   -2.050 &    12.0 &    14.1 &    16.0 \\
   25.00 &   4502.213 &    2.918 &   -0.340 \\
   25.00 &   4738.905 &    3.769 &   -4.770 \\
   25.00 &   4739.357 &    4.317 &   -2.970 \\
   25.00 &   5399.499 &    3.850 &   -0.290 \\
   25.00 &   5399.624 &    4.270 &   -4.490 \\
   25.00 &   5399.653 &    4.270 &   -4.290 \\
   25.00 &   6013.478 &    3.070 &   -0.869 \\
   25.00 &   6013.499 &    3.070 &   -1.081 \\
   25.00 &   6013.518 &    3.070 &   -1.354 \\
   25.00 &   6013.527 &    3.070 &   -1.558 \\
   25.00 &   6013.533 &    3.070 &   -1.764 \\
   25.00 &   6013.538 &    3.070 &   -1.412 \\
   25.00 &   6013.547 &    3.070 &   -1.433 \\
   25.00 &   6013.553 &    3.070 &   -1.588 \\
   25.00 &   6013.562 &    3.070 &   -1.910 \\
   25.00 &   6013.566 &    3.070 &   -1.956 \\
   25.00 &   6013.566 &    3.070 &   -2.513 \\
   25.00 &   6013.567 &    3.070 &   -2.132 \\
   25.00 &   6016.619 &    3.071 &   -1.541 \\
   25.00 &   6016.646 &    3.071 &   -1.378 \\
   25.00 &   6016.647 &    3.071 &   -0.763 \\
   25.00 &   6016.667 &    3.071 &   -1.373 \\
   25.00 &   6016.668 &    3.071 &   -1.026 \\
   25.00 &   6016.685 &    3.071 &   -1.357 \\
   25.00 &   6016.685 &    3.071 &   -1.475 \\
   25.00 &   6016.696 &    3.071 &   -1.541 \\
   25.00 &   6016.696 &    3.071 &   -1.804 \\
   25.00 &   6016.699 &    3.071 &   -1.737 \\
   25.00 &   6016.705 &    3.071 &   -2.503 \\
   25.00 &   6016.707 &    3.071 &   -1.378 \\
   25.00 &   6016.714 &    3.071 &   -1.373 \\
   25.00 &   6016.714 &    3.071 &   -1.737 \\
   25.00 &   6016.716 &    3.071 &   -1.475 \\
   26.00 &   4365.900 &    2.990 &   -2.250 &    45.1 &    40.7 &    51.3 \\
   26.00 &   4389.250 &    0.050 &   -4.580 &    67.2 &    62.4 &    71.9 \\
   26.00 &   4602.000 &    1.610 &   -3.150 &    65.0 &    60.9 &    71.6 \\
   26.00 &   4635.850 &    2.850 &   -2.340 &    49.4 &    44.2 &    56.9 \\
   26.00 &   4745.800 &    3.650 &   -1.270 &    69.2 &    63.9 &    77.6 \\
   26.00 &   4749.950 &    4.560 &   -1.240 &    28.1 &    24.8 &    36.0 \\
   26.00 &   4779.440 &    3.420 &   -2.160 &    33.6 &    28.7 &    40.2 \\
   26.00 &   4788.760 &    3.240 &   -1.730 &    60.2 &    55.2 &    67.8 \\
   26.00 &   4799.410 &    3.640 &   -2.130 &    27.8 &    23.7 &    35.5 \\
   26.00 &   4808.150 &    3.250 &   -2.690 &    20.2 &    16.2 &    26.5 \\
   26.00 &   4994.130 &    0.920 &   -3.080 &   100.7 &    92.9 &   104.7 \\
   26.00 &   5044.211 &    2.850 &   -2.060 &    67.9 &    60.6 &    73.1 \\
   26.00 &   5054.642 &    3.640 &   -1.920 &    30.6 &    26.1 &    38.8 \\
   26.00 &   5090.770 &    4.260 &   -0.490 &    83.3 &    76.8 &    92.7 \\
   26.00 &   5109.650 &    4.300 &   -0.730 &    66.2 &    59.5 &    73.5 \\
   26.00 &   5127.359 &    0.920 &   -3.310 &    91.5 &    85.0 &    97.3 \\
   26.00 &   5127.679 &    0.050 &   -6.120 &    15.7 &    10.1 &    18.4 \\
   26.00 &   5141.740 &    2.420 &   -2.230 &    78.7 &    73.1 &    87.2 \\
   26.00 &   5145.090 &    2.200 &   -3.080 &    47.0 &    39.8 &    54.9 \\
   26.00 &   5187.910 &    4.140 &   -1.260 &    46.5 &    42.2 &    56.2 \\
   26.00 &   5198.711 &    2.220 &   -2.140 &    91.1 &    83.8 &    96.6 \\
   26.00 &   5225.525 &    0.110 &   -4.790 &    65.7 &    58.5 &    71.2 \\
   26.00 &   5243.770 &    4.260 &   -0.990 &    53.3 &    48.7 &    62.7 \\
   26.00 &   5247.050 &    0.090 &   -4.960 &    59.2 &    53.2 &    65.6 \\
   26.00 &   5250.208 &    0.120 &   -4.940 &    59.2 &    52.0 &    65.4 \\
   26.00 &   5288.520 &    3.690 &   -1.510 &    48.6 &    44.4 &    57.8 \\
   26.00 &   5295.312 &    4.420 &   -1.590 &    23.3 &    18.7 &    28.6 \\
   26.00 &   5373.709 &    4.470 &   -0.740 &    55.7 &    50.4 &    62.4 \\
   26.00 &   5379.574 &    3.690 &   -1.510 &    52.7 &    48.3 &    59.1 \\
   26.00 &   5386.334 &    4.150 &   -1.670 &    25.1 &    20.5 &    31.5 \\
   26.00 &   5389.480 &    4.410 &   -0.450 &    76.0 &    70.9 &    84.2 \\
   26.00 &   5409.130 &    4.370 &   -1.060 &    46.8 &    41.3 &    54.7 \\
   26.00 &   5441.340 &    4.310 &   -1.630 &    23.4 &    21.3 &    33.2 \\
   26.00 &   5464.280 &    4.140 &   -1.580 &    28.1 &    25.3 &    37.2 \\
   26.00 &   5466.987 &    3.570 &   -2.230 &    26.6 &    22.3 &    34.8 \\
   26.00 &   5472.710 &    4.210 &   -1.520 &    33.6 &    28.7 &    43.4 \\
   26.00 &   5491.830 &    4.190 &   -2.190 &     9.3 &     7.1 &    14.1 \\
   26.00 &   5522.446 &    4.210 &   -1.310 &    35.3 &    31.0 &    44.6 \\
   26.00 &   5543.940 &    4.220 &   -1.040 &    54.7 &    49.6 &    61.5 \\
   26.00 &   5546.506 &    4.370 &   -1.180 &    41.7 &    38.4 &    50.9 \\
   26.00 &   5554.890 &    4.550 &   -0.360 &    84.4 &    77.9 &    97.7 \\
   26.00 &   5560.211 &    4.430 &   -1.090 &    43.2 &    39.4 &    52.8 \\
   26.00 &   5577.020 &    5.030 &   -1.460 &     7.9 &     6.6 &    10.9 \\
   26.00 &   5618.633 &    4.210 &   -1.270 &    41.7 &    37.2 &    50.8 \\
   26.00 &   5636.696 &    3.640 &   -2.560 &    14.5 &    11.7 &    18.9 \\
   26.00 &   5638.262 &    4.220 &   -0.770 &    69.2 &    64.2 &    78.4 \\
   26.00 &   5649.987 &    5.100 &   -0.800 &    28.2 &    24.2 &    36.0 \\
   26.00 &   5651.469 &    4.470 &   -1.750 &    13.2 &    11.4 &    17.6 \\
   26.00 &   5653.870 &    4.390 &   -1.540 &    29.4 &    26.3 &    35.7 \\
   26.00 &   5661.348 &    4.280 &   -1.760 &    16.4 &    12.8 &    21.9 \\
   26.00 &   5679.023 &    4.650 &   -0.750 &    49.7 &    45.8 &    58.0 \\
   26.00 &   5701.544 &    2.560 &   -2.160 &    77.1 &    69.6 &    82.2 \\
   26.00 &   5705.464 &    4.300 &   -1.360 &    30.9 &    26.2 &    38.1 \\
   26.00 &   5731.760 &    4.260 &   -1.200 &    49.8 &    44.3 &    56.8 \\
   26.00 &   5778.453 &    2.590 &   -3.440 &    17.8 &    13.4 &    21.9 \\
   26.00 &   5784.660 &    3.400 &   -2.530 &    19.5 &    15.7 &    27.8 \\
   26.00 &   5793.914 &    4.220 &   -1.620 &    27.1 &    21.7 &    33.2 \\
   26.00 &   5806.730 &    4.610 &   -0.950 &    44.8 &    40.0 &    54.1 \\
   26.00 &   5809.218 &    3.880 &   -1.610 &    40.2 &    34.6 &    50.2 \\
   26.00 &   5852.220 &    4.550 &   -1.230 &    31.6 &    27.8 &    40.4 \\
   26.00 &   5855.076 &    4.610 &   -1.480 &    15.9 &    13.0 &    23.0 \\
   26.00 &   5859.590 &    4.550 &   -0.580 &    63.4 &    58.4 &    72.5 \\
   26.00 &   5905.672 &    4.650 &   -0.690 &    48.6 &    44.4 &    58.6 \\
   26.00 &   5916.247 &    2.450 &   -2.940 &    46.7 &    39.8 &    54.2 \\
   26.00 &   5927.789 &    4.650 &   -1.040 &    32.9 &    30.4 &    41.9 \\
   26.00 &   5929.680 &    4.550 &   -1.310 &    31.2 &    28.0 &    39.4 \\
   26.00 &   5934.655 &    3.930 &   -1.070 &    67.8 &    61.0 &    75.6 \\
   26.00 &   5956.694 &    0.860 &   -4.610 &    42.6 &    36.6 &    52.6 \\
   26.00 &   6005.541 &    2.590 &   -3.430 &    15.9 &    12.6 &    22.1 \\
   26.00 &   6024.058 &    4.550 &   -0.020 &    97.6 &    92.4 &   110.1 \\
   26.00 &   6027.050 &    4.080 &   -1.090 &    56.1 &    51.8 &    63.2 \\
   26.00 &   6056.005 &    4.730 &   -0.400 &    63.9 &    59.4 &    71.5 \\
   26.00 &   6065.482 &    2.610 &   -1.530 &   109.1 &   101.5 &   117.7 \\
   26.00 &   6079.009 &    4.650 &   -1.020 &    37.1 &    32.4 &    46.5 \\
   26.00 &   6082.711 &    2.220 &   -3.570 &    27.9 &    22.3 &    34.5 \\
   26.00 &   6093.644 &    4.610 &   -1.300 &    24.7 &    19.7 &    31.3 \\
   26.00 &   6096.665 &    3.980 &   -1.810 &    30.0 &    24.6 &    36.5 \\
   26.00 &   6127.910 &    4.140 &   -1.400 &    39.7 &    36.1 &    50.5 \\
   26.00 &   6151.618 &    2.180 &   -3.280 &    42.0 &    36.4 &    50.2 \\
   26.00 &   6165.360 &    4.140 &   -1.460 &    36.8 &    31.7 &    45.6 \\
   26.00 &   6170.510 &    4.800 &   -0.380 &    67.8 &    62.0 &    80.0 \\
   26.00 &   6173.335 &    2.220 &   -2.880 &    61.1 &    55.1 &    68.5 \\
   26.00 &   6187.990 &    3.940 &   -1.620 &    38.9 &    33.7 &    48.0 \\
   26.00 &   6200.313 &    2.610 &   -2.420 &    65.6 &    58.6 &    71.2 \\
   26.00 &   6213.430 &    2.220 &   -2.520 &    76.3 &    69.5 &    82.9 \\
   26.00 &   6219.281 &    2.200 &   -2.430 &    82.3 &    76.3 &    88.7 \\
   26.00 &   6226.736 &    3.880 &   -2.100 &    21.8 &    18.4 &    30.2 \\
   26.00 &   6229.230 &    2.850 &   -2.830 &    29.8 &    26.1 &    40.3 \\
   26.00 &   6240.646 &    2.220 &   -3.290 &    39.7 &    34.7 &    50.2 \\
   26.00 &   6252.555 &    2.400 &   -1.690 &   112.1 &   103.5 &   120.6 \\
   26.00 &   6265.134 &    2.180 &   -2.550 &    78.5 &    72.8 &    86.7 \\
   26.00 &   6270.225 &    2.860 &   -2.540 &    43.2 &    38.4 &    51.9 \\
   26.00 &   6271.279 &    3.330 &   -2.700 &    19.1 &    14.3 &    23.1 \\
   26.00 &   6297.790 &    2.220 &   -2.710 &    66.9 &    61.6 &    75.5 \\
   26.00 &   6322.690 &    2.590 &   -2.430 &    66.9 &    61.9 &    76.0 \\
   26.00 &   6335.330 &    2.200 &   -2.260 &    89.9 &    82.7 &    95.4 \\
   26.00 &   6336.820 &    3.690 &   -0.930 &    99.4 &    89.4 &   105.4 \\
   26.00 &   6344.150 &    2.430 &   -2.920 &    51.0 &    44.3 &    59.3 \\
   26.00 &   6380.743 &    4.190 &   -1.320 &    43.6 &    38.6 &    52.9 \\
   26.00 &   6392.539 &    2.280 &   -4.030 &    13.0 &     9.0 &    16.8 \\
   26.00 &   6419.950 &    4.730 &   -0.240 &    74.7 &    68.3 &    85.0 \\
   26.00 &   6430.846 &    2.180 &   -2.010 &   104.9 &    96.0 &   112.1 \\
   26.00 &   6481.870 &    2.280 &   -2.980 &    55.8 &    49.3 &    64.6 \\
   26.00 &   6498.939 &    0.960 &   -4.700 &    37.1 &    31.3 &    46.8 \\
   26.00 &   6518.370 &    2.830 &   -2.450 &    47.7 &    42.8 &    56.8 \\
   26.00 &   6574.229 &    0.990 &   -5.010 &    22.8 &    16.6 &    29.0 \\
   26.00 &   6593.871 &    2.430 &   -2.390 &    75.5 &    70.2 &    84.2 \\
   26.00 &   6597.561 &    4.800 &   -0.970 &    34.8 &    31.9 &    43.3 \\
   26.00 &   6609.110 &    2.560 &   -2.680 &    58.1 &    50.9 &    65.3 \\
   26.00 &   6677.987 &    2.690 &   -1.420 &   117.7 &   107.4 &   123.6 \\
   26.00 &   6703.567 &    2.760 &   -3.020 &    28.7 &    24.8 &    36.3 \\
   26.00 &   6713.745 &    4.800 &   -1.400 &    15.4 &    12.5 &    20.1 \\
   26.00 &   6725.357 &    4.100 &   -2.190 &    12.2 &    10.5 &    18.2 \\
   26.00 &   6726.667 &    4.610 &   -1.030 &    37.8 &    34.0 &    48.5 \\
   26.00 &   6733.151 &    4.640 &   -1.470 &    19.0 &    16.8 &    25.9 \\
   26.00 &   6750.152 &    2.420 &   -2.620 &    67.3 &    61.0 &    72.9 \\
   26.00 &   6752.707 &    4.640 &   -1.200 &    27.8 &    23.3 &    36.5 \\
   26.00 &   6806.845 &    2.730 &   -3.110 &    27.0 &    21.3 &    33.1 \\
   26.00 &   6810.263 &    4.610 &   -0.990 &    41.6 &    35.9 &    51.2 \\
   26.00 &   6837.006 &    4.590 &   -1.690 &    13.6 &    10.9 &    19.0 \\
   26.00 &   6839.830 &    2.560 &   -3.350 &    23.1 &    17.6 &    31.1 \\
   26.00 &   6842.690 &    4.640 &   -1.220 &    29.7 &    24.7 &    37.3 \\
   26.00 &   6843.656 &    4.550 &   -0.830 &    51.2 &    46.4 &    59.7 \\
   26.00 &   6858.150 &    4.610 &   -0.940 &    42.4 &    38.3 &    51.3 \\
   26.10 &   4491.400 &    2.860 &   -2.660 &    67.5 &    68.3 &    76.3 \\
   26.10 &   4508.290 &    2.860 &   -2.520 &    77.5 &    79.9 &    84.4 \\
   26.10 &   4520.220 &    2.810 &   -2.650 &    72.8 &    74.1 &    78.2 \\
   26.10 &   4576.330 &    2.840 &   -2.950 &    55.2 &    56.1 &    62.8 \\
   26.10 &   4620.510 &    2.830 &   -3.210 &    42.5 &    44.3 &    50.8 \\
   26.10 &   4993.340 &    2.810 &   -3.730 &    29.5 &    31.1 &    38.5 \\
   26.10 &   5132.660 &    2.810 &   -4.170 &    15.5 &    16.5 &    24.1 \\
   26.10 &   5197.577 &    3.230 &   -2.220 &    72.8 &    74.7 &    80.4 \\
   26.10 &   5264.812 &    3.230 &   -3.130 &    35.1 &    38.3 &    45.0 \\
   26.10 &   5414.073 &    3.220 &   -3.580 &    19.4 &    20.0 &    25.1 \\
   26.10 &   5425.257 &    3.200 &   -3.220 &    31.6 &    33.4 &    41.0 \\
   26.10 &   6084.111 &    3.200 &   -3.830 &    14.4 &    15.9 &    19.9 \\
   26.10 &   6149.240 &    3.890 &   -2.750 &    27.7 &    29.7 &    35.9 \\
   26.10 &   6238.380 &    3.890 &   -2.630 &    34.2 &    35.4 &    44.0 \\
   26.10 &   6369.462 &    2.890 &   -4.110 &    13.3 &    13.9 &    18.2 \\
   26.10 &   6416.919 &    3.890 &   -2.750 &    31.9 &    33.4 &    40.2 \\
   26.10 &   6432.680 &    2.890 &   -3.570 &    32.7 &    33.5 &    40.1 \\
   26.10 &   6456.383 &    3.900 &   -2.050 &    53.2 &    55.8 &    61.4 \\
   27.00 &   4792.492 &    4.069 &   -2.360 \\
   27.00 &   4792.846 &    3.250 &   -0.070 \\
   27.00 &   4812.967 &    4.229 &   -3.750 \\
   27.00 &   4813.006 &    4.172 &   -3.440 \\
   27.00 &   4813.449 &    2.868 &   -2.120 \\
   27.00 &   4813.467 &    3.213 &    0.050 \\
   27.00 &   4813.794 &    2.872 &   -3.410 \\
   27.00 &   4813.966 &    3.295 &   -1.040 \\
   27.00 &   4814.042 &    2.628 &   -4.740 \\
   27.00 &   5212.691 &    3.512 &   -0.110 \\
   27.00 &   5213.316 &    4.471 &   -3.230 \\
   27.00 &   5301.039 &    1.709 &   -2.000 \\
   27.00 &   5301.101 &    2.135 &   -1.750 \\
   27.00 &   5483.344 &    1.709 &   -1.490 \\
   27.00 &   5483.374 &    4.501 &   -3.790 \\
   27.00 &   5483.955 &    3.629 &   -0.480 \\
   27.00 &   5647.107 &    4.146 &   -2.340 \\
   27.00 &   5647.234 &    2.278 &   -1.560 \\
   27.00 &   6093.143 &    1.739 &   -2.440 \\
   27.00 &   6454.990 &    3.629 &   -0.250 \\
   28.00 &   4831.180 &    3.610 &   -0.320 &    66.1 &    63.2 &    73.2 \\
   28.00 &   4866.270 &    3.540 &   -0.210 &    70.3 &    65.2 &    69.7 \\
   28.00 &   4904.420 &    3.540 &   -0.250 &    80.8 &    77.6 &    87.1 \\
   28.00 &   4913.980 &    3.740 &   -0.660 &    46.6 &    43.7 &    55.9 \\
   28.00 &   4946.040 &    3.800 &   -1.220 &    20.0 &    15.6 &    26.0 \\
   28.00 &   4952.290 &    3.610 &   -1.260 &    24.7 &    21.4 &    32.1 \\
   28.00 &   4953.208 &    3.740 &   -0.660 &    46.8 &    43.7 &    54.6 \\
   28.00 &   4976.135 &    3.610 &   -1.250 &    21.3 &    18.0 &    30.6 \\
   28.00 &   4998.220 &    3.610 &   -0.690 &    48.2 &    43.4 &    68.3 \\
   28.00 &   5010.938 &    3.640 &   -0.870 &    41.7 &    37.2 &    47.9 \\
   28.00 &   5082.350 &    3.660 &   -0.590 &    59.1 &    53.7 &    66.4 \\
   28.00 &   5084.110 &    3.680 &   -0.060 &    83.6 &    78.9 &    88.0 \\
   28.00 &   5088.960 &    3.680 &   -1.290 &    21.4 &    17.5 &    24.6 \\
   28.00 &   5094.420 &    3.830 &   -1.070 &    22.9 &    19.8 &    28.2 \\
   28.00 &   5102.970 &    1.680 &   -2.660 &    41.2 &    36.5 &    49.2 \\
   28.00 &   5157.980 &    3.610 &   -1.510 &    13.1 &    12.2 &    18.6 \\
   28.00 &   5176.560 &    3.900 &   -0.440 &    46.8 &    44.0 &    54.5 \\
   28.00 &   5392.330 &    4.150 &   -1.320 &     9.5 &     7.9 &    12.5 \\
   28.00 &   5578.729 &    1.680 &   -2.570 &    47.1 &    41.5 &    55.0 \\
   28.00 &   5587.870 &    1.930 &   -2.440 &    44.6 &    40.0 &    53.5 \\
   28.00 &   5589.358 &    3.900 &   -1.140 &    19.8 &    18.0 &    25.9 \\
   28.00 &   5593.746 &    3.900 &   -0.780 &    33.4 &    30.4 &    43.5 \\
   28.00 &   5625.320 &    4.090 &   -0.730 &    29.4 &    26.4 &    36.3 \\
   28.00 &   5628.350 &    4.090 &   -1.320 &    10.9 &     8.7 &    14.6 \\
   28.00 &   5638.750 &    3.900 &   -1.700 &     5.3 &     4.3 &    10.4 \\
   28.00 &   5641.880 &    4.110 &   -1.020 &    17.6 &    14.2 &    21.6 \\
   28.00 &   5643.078 &    4.160 &   -1.250 &    10.9 &     8.9 &    14.5 \\
   28.00 &   5694.990 &    4.090 &   -0.630 &    34.3 &    30.2 &    42.6 \\
   28.00 &   5748.360 &    1.680 &   -3.240 &    20.6 &    17.4 &    28.2 \\
   28.00 &   5754.670 &    1.930 &   -1.850 &    65.4 &    61.2 &    73.6 \\
   28.00 &   5805.217 &    4.170 &   -0.640 &    33.0 &    28.5 &    40.0 \\
   28.00 &   5847.010 &    1.676 &   -3.410 &    17.1 &    13.2 &    22.5 \\
   28.00 &   5996.740 &    4.236 &   -1.010 &    13.8 &    12.9 &    20.4 \\
   28.00 &   6007.317 &    1.677 &   -3.410 &    17.7 &    15.6 &    25.1 \\
   28.00 &   6086.282 &    4.270 &   -0.510 &    34.0 &    28.9 &    43.8 \\
   28.00 &   6108.116 &    1.680 &   -2.440 &    53.2 &    48.4 &    66.8 \\
   28.00 &   6111.080 &    4.088 &   -0.810 &    26.3 &    23.6 &    34.1 \\
   28.00 &   6119.760 &    4.270 &   -1.316 &     7.5 &     6.2 &    10.5 \\
   28.00 &   6128.984 &    1.677 &   -3.360 &    19.1 &    16.1 &    26.2 \\
   28.00 &   6130.135 &    4.270 &   -0.960 &    14.2 &    12.2 &    21.9 \\
   28.00 &   6175.370 &    4.089 &   -0.550 &    39.5 &    37.0 &    49.5 \\
   28.00 &   6176.811 &    4.090 &   -0.260 &    53.7 &    50.3 &    65.4 \\
   28.00 &   6177.242 &    1.830 &   -3.510 &    10.2 &     9.4 &    15.3 \\
   28.00 &   6186.717 &    4.110 &   -0.960 &    23.6 &    18.4 &    31.3 \\
   28.00 &   6204.604 &    4.090 &   -1.140 &    16.3 &    13.9 &    22.7 \\
   28.00 &   6223.971 &    4.105 &   -1.466 &    20.9 &    17.3 &    29.2 \\
   28.00 &   6230.100 &    4.110 &   -1.132 &    13.8 &    11.9 &    21.2 \\
   28.00 &   6322.169 &    4.154 &   -1.210 &    12.2 &     9.1 &    18.0 \\
   28.00 &   6327.600 &    1.680 &   -3.060 &    28.2 &    22.7 &    37.9 \\
   28.00 &   6360.810 &    4.170 &   -1.150 &    12.7 &    12.6 &    16.5 \\
   28.00 &   6378.233 &    4.154 &   -1.386 &    22.0 &    20.8 &    31.6 \\
   28.00 &   6482.810 &    1.930 &   -2.760 &    50.8 &    49.0 &    39.6 \\
   28.00 &   6598.611 &    4.236 &   -0.910 &    17.0 &    15.6 &    25.1 \\
   28.00 &   6635.130 &    4.420 &   -0.720 &    17.3 &    14.8 &    23.8 \\
   28.00 &   6643.630 &    1.680 &   -2.000 &    84.5 &    77.5 &    94.8 \\
   28.00 &   6767.772 &    1.830 &   -2.170 &    71.1 &    66.1 &    80.2 \\
   28.00 &   6772.315 &    3.660 &   -0.990 &    37.9 &    36.0 &    50.9 \\
   28.00 &   6842.043 &    3.658 &   -1.500 &    19.1 &    17.0 &    23.4 \\
   29.00 &   5218.197 &    3.814 &    0.480 \\
   30.00 &   4722.159 &    4.030 &   -0.380 &    64.6 &    65.2 &    69.6 \\
   30.00 &   4810.534 &    4.080 &   -0.160 &    67.7 &    68.2 &    74.3 \\
   30.00 &   6362.350 &    5.790 &    0.140 &    14.9 &    16.0 &    19.8 \\
   38.00 &   4607.338 &    0.000 &    0.283 &    37.2 &    29.8 &    45.7 \\
   38.10 &   4161.800 &    2.940 &   -0.500 &    19.4 &    19.0 &    25.7 \\
   39.10 &   4854.867 &    0.992 &   -0.380 &    36.4 &    35.3 &    43.8 \\
   39.10 &   4883.685 &    1.084 &    0.070 &    48.7 &    48.0 &    56.3 \\
   39.10 &   4900.110 &    1.033 &   -0.090 &    44.0 &    45.0 &    54.6 \\
   39.10 &   5200.413 &    0.992 &   -0.570 &    26.9 &    27.0 &    36.7 \\
   40.10 &   4050.320 &    0.713 &   -1.060 &    18.5 &    14.5 &    22.3 \\
   40.10 &   4208.980 &    0.713 &   -0.510 &    37.0 &    38.0 &    42.9 \\
   56.10 &   5853.696 &    0.604 &   -2.915 &    56.4 &    54.8 &    63.0 \\
   56.10 &   6141.697 &    0.704 &   -2.495 &   112.1 &   104.3 &   115.6 \\
   56.10 &   6496.900 &    0.604 &   -2.000 &    94.1 &    93.3 &    98.7 \\
   56.10 &   5853.686 &    0.604 &   -2.066 \\
   56.10 &   5853.687 &    0.604 &   -2.066 \\
   56.10 &   5853.687 &    0.604 &   -2.009 \\
   56.10 &   5853.688 &    0.604 &   -2.009 \\
   56.10 &   5853.689 &    0.604 &   -2.215 \\
   56.10 &   5853.689 &    0.604 &   -2.215 \\
   56.10 &   5853.690 &    0.604 &   -1.010 \\
   56.10 &   5853.690 &    0.604 &   -1.466 \\
   56.10 &   5853.690 &    0.604 &   -1.914 \\
   56.10 &   5853.690 &    0.604 &   -2.620 \\
   56.10 &   5853.690 &    0.604 &   -1.010 \\
   56.10 &   5853.690 &    0.604 &   -1.466 \\
   56.10 &   5853.690 &    0.604 &   -1.914 \\
   56.10 &   5853.690 &    0.604 &   -2.620 \\
   56.10 &   5853.690 &    0.604 &   -1.010 \\
   56.10 &   5853.691 &    0.604 &   -2.215 \\
   56.10 &   5853.692 &    0.604 &   -2.215 \\
   56.10 &   5853.693 &    0.604 &   -2.009 \\
   56.10 &   5853.693 &    0.604 &   -2.009 \\
   56.10 &   5853.694 &    0.604 &   -2.066 \\
   56.10 &   5853.694 &    0.604 &   -2.066 \\
   56.10 &   6141.725 &    0.704 &   -2.456 \\
   56.10 &   6141.725 &    0.704 &   -2.456 \\
   56.10 &   6141.727 &    0.704 &   -1.311 \\
   56.10 &   6141.727 &    0.704 &   -1.311 \\
   56.10 &   6141.728 &    0.704 &   -2.284 \\
   56.10 &   6141.728 &    0.704 &   -2.284 \\
   56.10 &   6141.729 &    0.704 &   -0.503 \\
   56.10 &   6141.729 &    0.704 &   -1.214 \\
   56.10 &   6141.729 &    0.704 &   -0.503 \\
   56.10 &   6141.729 &    0.704 &   -1.214 \\
   56.10 &   6141.730 &    0.704 &   -0.077 \\
   56.10 &   6141.730 &    0.704 &   -0.077 \\
   56.10 &   6141.730 &    0.704 &   -0.077 \\
   56.10 &   6141.731 &    0.704 &   -0.709 \\
   56.10 &   6141.731 &    0.704 &   -1.327 \\
   56.10 &   6141.731 &    0.704 &   -0.709 \\
   56.10 &   6141.731 &    0.704 &   -1.327 \\
   56.10 &   6141.732 &    0.704 &   -0.959 \\
   56.10 &   6141.732 &    0.704 &   -1.281 \\
   56.10 &   6141.732 &    0.704 &   -0.959 \\
   56.10 &   6141.733 &    0.704 &   -1.281 \\
   56.10 &   6496.898 &    0.604 &   -1.886 \\
   56.10 &   6496.899 &    0.604 &   -1.886 \\
   56.10 &   6496.901 &    0.604 &   -1.186 \\
   56.10 &   6496.902 &    0.604 &   -1.186 \\
   56.10 &   6496.906 &    0.604 &   -0.739 \\
   56.10 &   6496.906 &    0.604 &   -0.739 \\
   56.10 &   6496.910 &    0.604 &   -0.380 \\
   56.10 &   6496.910 &    0.604 &   -0.380 \\
   56.10 &   6496.910 &    0.604 &   -0.380 \\
   56.10 &   6496.916 &    0.604 &   -1.583 \\
   56.10 &   6496.916 &    0.604 &   -1.583 \\
   56.10 &   6496.917 &    0.604 &   -1.186 \\
   56.10 &   6496.918 &    0.604 &   -1.186 \\
   56.10 &   6496.920 &    0.604 &   -1.186 \\
   56.10 &   6496.922 &    0.604 &   -1.186 \\
   57.10 &   4662.500 &    0.000 &   -1.240 &     7.3 &     6.2 &     5.7 \\
   58.10 &   4042.581 &    0.495 &    0.000 &     7.8 &     7.9 &    11.0 \\
   58.10 &   4073.474 &    0.477 &    0.210 &    13.7 &    12.8 &    39.5 \\
   58.10 &   4364.653 &    0.495 &   -0.170 &     8.8 &     9.2 &    11.5 \\
   58.10 &   4523.075 &    0.516 &   -0.080 &    11.2 &    10.5 &    13.8 \\
   58.10 &   4562.359 &    0.477 &    0.210 &    20.0 &    18.4 &    22.2 \\
   58.10 &   5274.229 &    1.044 &    0.130 &     5.9 &     5.7 &     7.5 \\
   60.10 &   4021.330 &    0.320 &   -0.100 &    10.5 &    10.8 &     5.7 \\
   60.10 &   4446.380 &    0.204 &   -0.350 &     7.7 &     7.8 &     5.6 \\
   63.10 &   4129.720 &    0.000 &    0.220 &    46.2 &    46.7 &    55.5 \\
   64.10 &   4251.731 &    0.382 &   -0.220 &     8.8 &     7.7 &    14.0 \\
   66.10 &   4077.970 &    0.103 &   -0.040 &    31.6 &    26.0 &    27.4 \\
   66.10 &   4103.310 &    0.103 &   -0.380 &     9.8 &    10.4 &    13.1 \\
   66.10 &   4449.700 &    0.000 &   -1.030 &     5.0 &     4.1 &     3.8 \\
\label{linelist}
\end{longtable}
}

\onltab{2}{
\setcounter{table}{2}
\begin{table*}
\centering
\caption{Differential abundances for the stars $\zeta^{1}$ Ret and $\zeta^{2}$ Ret relative to the Sun,
and $\zeta^{1}$ Ret relative to $\zeta^{2}$ Ret.
We also present the observational errors $\sigma_{obs}$, errors due to stellar
parameters $\sigma_{par}$, as well as the total error $\sigma_{TOT}$.}
%\hskip -0.35in
%\scriptsize
\begin{tabular}{ccccccccccccc}
\hline
\hline
  & \multicolumn{4}{c}{($\zeta^{1}$ Ret - Sun)} &  \multicolumn{4}{c}{($\zeta^{2}$ Ret - Sun)} & \multicolumn{4}{c}{($\zeta^{1}$ Ret- $\zeta^{2}$ Ret)}\\
Element & [X/Fe] & $\sigma_{obs}$ & $\sigma_{par}$ & $\sigma_{TOT}$ & [X/Fe]  & $\sigma_{obs}$ & $\sigma_{par}$ & $\sigma_{TOT}$ & [X/Fe]  & $\sigma_{obs}$ & $\sigma_{par}$ & $\sigma_{TOT}$  \\
\hline
 {[C I/Fe]}  & 0.050 & 0.060 & 0.047 & 0.076 & 0.051 & 0.060 & 0.021 & 0.064 & 0.000 & 0.040 & 0.018 & 0.044 \\
 {[O I/Fe]}  & -0.045 & 0.049 & 0.023 & 0.054 & 0.095 & 0.054 & 0.020 & 0.057 & -0.139 & 0.005 & 0.016 & 0.017 \\
 {[Na I/Fe]}  & 0.004 & 0.040 & 0.010 & 0.041 & 0.009 & 0.016 & 0.009 & 0.018 & -0.003 & 0.026 & 0.007 & 0.027 \\
 {[Mg I/Fe]}  & 0.110 & 0.044 & 0.019 & 0.048 & 0.133 & 0.039 & 0.010 & 0.040 & -0.022 & 0.011 & 0.008 & 0.014 \\
 {[Al I/Fe]}  & 0.060 & 0.028 & 0.011 & 0.030 & 0.007 & 0.058 & 0.010 & 0.059 & 0.054 & 0.034 & 0.008 & 0.035 \\
 {[Si I/Fe]}  & 0.024 & 0.012 & 0.002 & 0.014 & 0.045 & 0.012 & 0.002 & 0.013 & -0.020 & 0.005 & 0.001 & 0.006 \\
 {[S I/Fe]}  & 0.096 & 0.103 & 0.021 & 0.105 & 0.176 & 0.048 & 0.017 & 0.051 & -0.080 & 0.054 & 0.014 & 0.056 \\
 {[Ca I/Fe]}  & 0.074 & 0.008 & 0.011 & 0.015 & 0.067 & 0.011 & 0.006 & 0.012 & 0.008 & 0.006 & 0.005 & 0.008 \\
 {[Sc I/Fe]}  & 0.083 & 0.070 & 0.028 & 0.076 & 0.075 & 0.003 & 0.026 & 0.026 & 0.009 & 0.073 & 0.021 & 0.076 \\
 {[Sc II/Fe]}  & 0.003 & 0.017 & 0.025 & 0.031 & -0.011 & 0.018 & 0.011 & 0.021 & 0.015 & 0.002 & 0.008 & 0.009 \\
 {[Ti I/Fe]}  & 0.085 & 0.008 & 0.008 & 0.012 & 0.070 & 0.009 & 0.005 & 0.011 & 0.016 & 0.006 & 0.004 & 0.007 \\
 {[Ti II/Fe]}  & 0.051 & 0.014 & 0.016 & 0.021 & 0.060 & 0.011 & 0.006 & 0.013 & -0.008 & 0.008 & 0.005 & 0.010 \\
 {[V I/Fe]}  & -0.236 & 0.016 & 0.014 & 0.022 & -0.278 & 0.017 & 0.006 & 0.019 & 0.044 & 0.008 & 0.005 & 0.010 \\
 {[Cr I/Fe]}  & 0.031 & 0.010 & 0.009 & 0.014 & 0.034 & 0.012 & 0.005 & 0.013 & -0.002 & 0.008 & 0.004 & 0.010 \\
 {[Cr II/Fe]}  & 0.047 & 0.016 & 0.022 & 0.028 & 0.045 & 0.016 & 0.009 & 0.019 & 0.003 & 0.022 & 0.007 & 0.023 \\
 {[Mn I/Fe]}  & -0.369 & 0.067 & 0.024 & 0.071 & -0.285 & 0.022 & 0.011 & 0.024 & 0.043 & 0.023 & 0.009 & 0.024 \\
 {[Co I/Fe]}  & -0.155 & 0.011 & 0.018 & 0.022 & -0.173 & 0.010 & 0.008 & 0.013 & 0.018 & 0.015 & 0.007 & 0.017 \\
 {[Ni I/Fe]}  & -0.006 & 0.010 & 0.004 & 0.012 & 0.005 & 0.011 & 0.002 & 0.012 & -0.010 & 0.005 & 0.002 & 0.006 \\
 {[Cu I/Fe]}  & -0.185 & 0.060 & 0.047 & 0.076 & -0.215 & 0.060 & 0.021 & 0.064 & 0.031 & 0.040 & 0.018 & 0.044 \\
 {[Zn I/Fe]}  & 0.041 & 0.013 & 0.044 & 0.046 & 0.068 & 0.013 & 0.010 & 0.016 & -0.025 & 0.010 & 0.010 & 0.015 \\
 {[Sr I/Fe]}  & -0.053 & 0.060 & 0.078 & 0.099 & -0.110 & 0.060 & 0.031 & 0.067 & 0.058 & 0.040 & 0.030 & 0.050 \\
 {[Sr II/Fe]}  & 0.008 & 0.060 & 0.032 & 0.068 & -0.005 & 0.060 & 0.017 & 0.062 & 0.014 & 0.040 & 0.013 & 0.042 \\
 {[Y II/Fe]}  & -0.050 & 0.026 & 0.049 & 0.056 & -0.057 & 0.012 & 0.014 & 0.019 & 0.007 & 0.017 & 0.014 & 0.022 \\
 {[Zr II/Fe]}  & 0.042 & 0.025 & 0.060 & 0.066 & 0.003 & 0.074 & 0.022 & 0.077 & 0.040 & 0.099 & 0.019 & 0.101 \\
 {[Ba II/Fe]}  & -0.095 & 0.065 & 0.033 & 0.073 & -0.208 & 0.057 & 0.015 & 0.059 & 0.114 & 0.022 & 0.013 & 0.026 \\
 {[La II/Fe]}  & 0.287 & 0.060 & 0.027 & 0.066 & 0.261 & 0.060 & 0.023 & 0.064 & 0.027 & 0.040 & 0.015 & 0.043 \\
 {[Ce II/Fe]}  & -0.073 & 0.134 & 0.014 & 0.134 & -0.047 & 0.137 & 0.009 & 0.137 & -0.035 & 0.016 & 0.006 & 0.018 \\
 {[Nd II/Fe]}  & 0.399 & 0.112 & 0.028 & 0.116 & 0.468 & 0.115 & 0.022 & 0.117 & -0.068 & 0.003 & 0.015 & 0.016 \\
 {[Eu II/Fe]}  & -0.110 & 0.060 & 0.200 & 0.209 & -0.133 & 0.060 & 0.040 & 0.072 & 0.024 & 0.040 & 0.054 & 0.067 \\
 {[Gd II/Fe]}  & -0.079 & 0.060 & 0.027 & 0.066 & -0.100 & 0.060 & 0.022 & 0.064 & 0.022 & 0.040 & 0.014 & 0.043 \\
 {[Dy II/Fe]}  & 0.211 & 0.124 & 0.035 & 0.129 & 0.174 & 0.057 & 0.016 & 0.060 & 0.037 & 0.088 & 0.014 & 0.090 \\
\hline
\end{tabular}
\normalsize
%\tablebib{R1 \citep{abt79}, R2 \citep{mermilliod83}, R3 \citep{shore87}}
\label{table.abunds}
\end{table*}
}

\end{document}